\documentclass[reprint,superscriptaddress,preprintnumbers,amsmath,amssymb,aps,prl,nofootinbib]{revtex4-1}

\usepackage{placeins} 
\usepackage[normalem]{ulem} 
\usepackage{graphicx}  
\usepackage{xcolor}
\usepackage{dcolumn}   
\usepackage{bm}        
\usepackage{hyperref}  
\usepackage{multirow}
\usepackage{siunitx}
\graphicspath{
{./}
  {./media/}
}

\def\al{\alpha} 
\def\be{\beta} 

\def\de{\delta}

\def\ze{\zeta}

\def\ka{\kappa}

\newcommand{\ben}{\begin{equation}}
\newcommand{\een}{\end{equation}}
\newcommand{\bea}{\begin{eqnarray}}
\newcommand{\eea}{\end{eqnarray}}
\newcommand{\ba}{\begin{array}}
\newcommand{\ea}{\end{array}}
\newcommand{\bal}{\begin{align}}
\newcommand{\eal}{\end{align}}
\newcommand{\bit}{\begin{itemize}}
\newcommand{\eit}{\end{itemize}}

\newcommand{\ie}{\textit{i.e.\ }}

\newcommand{\fa}{f_\text{a}} 
\newcommand{\StrWid}{r_\text{s}} 
\newcommand{\tOff}{t_0}

\newcommand{\zeEst}{\hat{\zeta}}

\newcommand{\zzMean}{1.19}
\newcommand{\zzErr}{0.20}
\newcommand{\LogCorMean}{-0.04}
\newcommand{\LogCorErr}{0.30}

\newcommand{\mysection}[1]{{\it #1: }}

\begin{document}

\preprint{HIP-2020-1/TH}

\title{The scaling density of axion strings}

\newcommand{\Sussex}{\affiliation{
Department of Physics and Astronomy,
University of Sussex, Falmer, Brighton BN1 9QH,
U.K.}}

\newcommand{\HIPetc}{\affiliation{
Department of Physics and Helsinki Institute of Physics,
PL 64, 
FI-00014 University of Helsinki,
Finland
}}

\newcommand{\EHU}{\affiliation{
Department of Theoretical Physics,
University of the Basque Country UPV/EHU, 
48080 Bilbao,
Spain
}}

\author{Mark Hindmarsh}
\email{mark.hindmarsh@helsinki.fi}
\HIPetc
\Sussex

\author{Joanes Lizarraga}
\email{joanes.lizarraga@ehu.eus}
\EHU

\author{Asier Lopez-Eiguren}
\email{asier.lopezeiguren@helsinki.fi}
\HIPetc

\author{Jon Urrestilla}
\email{jon.urrestilla@ehu.eus}
\EHU

\date{\today}

\begin{abstract}
In the QCD axion dark matter scenario with post-inflationary Peccei-Quinn symmetry breaking, the number density of axions, and hence the dark matter density, depends on the length of string per unit volume at cosmic time $t$, by convention written $\zeta/t^2$. The expectation has been that the dimensionless parameter $\zeta$ tends to a constant $\zeta_0$, a feature of a string network known as scaling. It has recently been claimed that in larger numerical simulations $\zeta$ shows a logarithmic increase with time, while theoretical modelling suggests an inverse logarithmic correction. Either case would result in a large enhancement of the string density at the QCD transition, and a substantial revision to the axion mass required for the axion to constitute all of the dark matter. With a set of new simulations of global strings we compare the standard scaling (constant-$\zeta$) model to the logarithmic growth and inverse-logarithmic correction models. In the standard scaling model, by fitting to linear growth in the mean string separation $\xi = t/\sqrt{\zeta}$, we find $\zeta_0 = \zzMean \pm \zzErr$. We conclude that the apparent corrections to $\zeta$ are artefacts of the initial conditions, rather than a property of the scaling network. The residuals from the constant-$\zeta$ (linear $\xi$) fit also show no evidence for logarithmic growth, restoring confidence that numerical simulations can be simply extrapolated from the Peccei-Quinn symmetry-breaking scale to the QCD scale. Re-analysis of previous work on the axion number density suggests that recent estimates of the axion dark matter mass in the post-inflationary symmetry-breaking scenario we study should be increased by about 50\%. 

\end{abstract}

\maketitle

\mysection{Introduction}
The Peccei-Quinn (PQ) mechanism, which solves the strong CP problem of QCD 
by extending the Standard Model with an extra U(1) global symmetry 
\cite{Peccei:1977hh,*Peccei:1977ur}, 
brings with it a long-lived pseudoscalar particle, the axion \cite{Weinberg:1977ma,*Wilczek:1977pj}. 
A universe where light axions \cite{Kim:1979if,*Shifman:1979if,Zhitnitsky:1980tq,*Dine:1981rt}  
constitute the dark matter \cite{Preskill:1982cy,*Abbott:1982af,*Dine:1982ah} 
is one of the most promising scenarios in the current cosmological paradigm. 

If the PQ symmetry is spontaneously broken after primordial inflation, 
axion strings are formed \cite{Davis:1986xc},
a variety of global cosmic string \cite{Hindmarsh:1994re,Vilenkin:2000jqa}. 
They survive until the QCD confinement transition, when they become connected by domain walls
made of the CP-odd gluon condensate  \cite{Sikivie:1982qv,Georgi:1982ph}, 
and are annihilated.  
Most of the energy is left behind in the form of axion radiation, produced through the lifetime of the 
string network and during the annihilation phase. 
The axion radiation can also be viewed as light massive particles, whose 
number density depends on the length of string per unit volume $\zeta/t^2$, where $t$ is cosmic time. 
The important dimensionless parameter $\ze$ can be established only by numerical simulations. 

The usual expectation 
(see \cite{Shellard:1998mi,Sikivie:2006ni,Kim:2008hd})
is that the string density parameter $\zeta$ converges to a constant  within a few Hubble times after the 
network is formed, part of a wider assumption known as scaling. 
Scaling means that the string network is statistically self-similar; 
\ie all macroscopic quantities with the dimensions of length and time are proportional to the Hubble length and time. 
Earlier simulations of global cosmic strings 
\cite{Yamaguchi:1998gx,Yamaguchi:1999wt,Yamaguchi:2000fg,Yamaguchi:2002sh,
Hiramatsu:2010yu,Hiramatsu:2012gg,Kawasaki:2014sqa,Lopez-Eiguren:2017dmc}
were consistent with scaling with $\zeta \sim 1$,
and there is good theoretical understanding of scaling from 
modelling the global properties of the network \cite{Martins:1996jp,Martins:2000cs}.

However, several groups have recently claimed that 
$\zeta$ shows a logarithmic increase with time 
\cite{Gorghetto:2018myk,Kawasaki:2018bzv,Vaquero:2018tib,Buschmann:2019icd}.
An argument for expecting a scaling violation is based on the logarithmic growth in the effective 
string tension of a global string with their mean separation. 
If there is no corresponding change in the energy loss rate per unit length, 
the string length density parameter should grow \cite{Fleury:2015aca,Klaer:2017qhr,Klaer:2017ond,Hill:1987bw}. 

In fact this argument does not lead to
logarithmic growth of $\zeta$; instead it gives a leading correction to scaling of an inverse logarithm \cite{Martins:2018dqg}.
Nonetheless, either behaviour would lead to a larger asymptotic string density parameter, 
which would lead to an increase of the axion number density, 
and hence a decrease in the axion mass required to match the current dark matter mass density. 

In this work we present results from a new set of numerical simulations of global strings.  
We explore the effect of different initial string densities and lattice sizes. 
We compare the results for the string density in three different 
two-parameter models defined below: standard scaling,
logarithmic, and inverse-logarithmic.
We demonstrate that all simulations are consistent with 
standard scaling, and determine the asymptotic 
string length density parameter $\zeta_0$ to the best precision to date. 

We conclude that the axion string density shows excellent scaling 
following the PQ phase transition, justifying a constant-$\zeta$ extrapolation to the QCD transition.  
We re-examine previous results to see how estimates of the axion number density, and 
hence the axion dark matter mass, are affected. 

\mysection{Model and Simulations}
\label{sec:Model&Sims}
The simplest axion models \cite{Kim:1979if,*Shifman:1979if,Zhitnitsky:1980tq,*Dine:1981rt} 
break the U(1)$_{\rm PQ}$ symmetry with a scalar gauge singlet field, 
which we can write as a 
real scalar doublet $\Phi$ with action 
\ben
S=\int d^4 x \sqrt{-g} \Big( \frac{1}{2} \partial_{\mu} \Phi \partial^{\mu}\Phi- \frac{1}{4}\lambda(\Phi^2-\eta^2)^2 \Big),
\label{eq:ac}
\een
where $\lambda$ is the self-coupling of the scalar field and $\eta$ its vacuum expectation value.  
The metric $g_{\mu\nu}$ is the spatially flat Friedmann-Lema\^itre-Robertson-Walker metric 
with comoving spatial coordinates $ds^2=dt^2-a^2(t)d\bm{r}^2$, where $a(t)$ is the scale factor and $t$ is physical time.

When PQ symmetry is spontaneously broken, axion strings are formed and 
one massless 
Goldstone boson and one massive boson arise. Even though the axions acquire a small mass, when the coupling to QCD fields are considered \cite{Peccei:1977hh,Peccei:1977ur}, at high temperatures the axion mass can be neglected, and the field 
obeys the following dynamics: 

\ben
{\Phi}''+2\frac{{a'}}{a}{\Phi'}-\nabla^2 \Phi = -a^2 \lambda (\Phi^2-\eta^2)\Phi,
\label{eq:eom}
\een
where the primes represent derivatives with respect to the conformal time $\tau = \int dt a^{-1}(t)$. 
For axion string evolution,  $a \propto \tau$.

The evolution of the field is simulated 
with a discretised version of Eq.~(\ref{eq:eom}), 
parallelised using the LATfield2 library \cite{LatField2d}. 
We use  cubic lattices with periodic boundary conditions, which impose an upper limit in the dynamical range of the simulation of half a light-crossing time, 
beyond which it is possible for the Goldstone modes to show finite volume effects in their propagation.  
Note that we do not use the Press-Ryden-Spergel method \cite{Press:1989yh}; 
data is taken while the string core has constant physical width $\StrWid = m_\text{s}^{-1}$ 
and shrinking comoving width.

We use initial conditions designed to drive the system quickly to scaling. 
To this end, a satisfactory initial field configuration is given by the scalar field velocities $\dot\Phi$ 
set to zero and the components of $\Phi$ 
to be Gaussian random fields with power spectrum,
$P_{\Phi}(\mathbf{k})={A}\left[{1+(k\ell_{\phi})^2}\right]^{-1}$,
with $A$ chosen so that $\langle \Phi^2 \rangle=\eta^2$. 
We use comoving correlation lengths $\ell_{\phi}\eta = (5, 10, 20)$. 
We run with lattice sites per side $N = [1k, 2k, 4k]$ (where $k = 1024$), and 
perform 4 independent runs in each different lattice and for each correlation length.

In order to 
remove energy from the initial configuration, 
 $\lambda$ is time-dependent in the preparation phase, 
so that we can arrange $m_\text{s} a \simeq 2\eta$ 
at $\tau_{\rm ini}\eta=50$, and apply a period of diffusive evolution with unit diffusion constant,  
until $\tau_{\rm diff}\eta=70$.    We then apply the  
second order equations (\ref{eq:eom}), 
allowing the comoving width of the strings to grow to their physical value   
at $\tau_{\rm cg}\eta=[144.9,196.2,271.1]$ for $N = [1k, 2k, 4k]$. 

The physical evolution begins at $\tau_{\rm cg}$ and ends at $\tau_{\rm end}\eta=[300,550,1050]$, 
when $m_s a = 2\eta$, during which $\lambda=2$ is constant. 
We normalise the scale factor so that $a(\tau_{\rm end}) = 1$.   
The comoving lattice spacing is $\delta x\eta = 0.5$, the conformal timestep during diffusion is $\delta \tau= \de x/30$ 
and during second order evolution $\de\tau = \de x/5$. In the subsequent figures and tables the unit of length is $\eta^{-1}$.


\mysection{Measurements and results}
The evolution of the string network can be tracked by the mean string separation $\xi$, 
defined in terms of the mean string length $\ell_{\rm s}$ in the simulation volume $\mathcal{V}$ as 
\ben
\label{e:xisDef}
\xi=\sqrt{{\mathcal{V}}/{\ell_{\rm s}}}.
\een
The physical string length $\ell_s$ is the number of plaquettes pierced by strings multiplied by the 
physical lattice spacing $a\delta x$,  
corrected by factor of $2/3$ to compensate for the Manhattan effect \cite{Fleury:2015aca}. 
Such plaquettes are identified calculating the ``winding'' of the phase of the field around each plaquette 
of the lattice \cite{Vachaspati:1984dz}. 

A dimensionless measure of the length of string per unit volume 
\cite{Vilenkin:2000jqa,Klaer:2017ond,Gorghetto:2018myk,Vaquero:2018tib,Kawasaki:2018bzv,Martins:2018dqg} is 
\ben
\zeta = {\ell_{\rm s} t^2 }/{\mathcal{V}} = {t^2}/{\xi^2},
\label{eq:zeta}
\een
which in a radiation-dominated universe 
is four times the number of Hubble lengths of string per 
Hubble volume (note that some authors use $\xi$ to denote this quantity).

As there is no fixed length scale in the string equations of motion, 
string networks are expected to evolve towards a self-similar or scaling regime, in which the only length scale is $t$ 
\cite{Hindmarsh:1994re,Vilenkin:2000jqa,Martins:1996jp}. 
Hence $\xi$ should increase linearly with time, and $\zeta$ should evolve towards a constant.
However, the formation and initial evolution of the network introduces a time scale, which can be taken to 
be the $t$-axis intercept of a linear fit to $\xi(t)$ \cite{Bevis:2010gj}. 
We call this the initial string evolution parameter, and denote it $\tOff$.
Over cosmological timescales the ratio $\tOff/t \to 0$; however, in numerical simulations 
it must be taken into account when extracting the scaling value of $\ze$, which we denote $\ze_0$.

 \begin{figure}[htbp]
    \centering
    \includegraphics[width=3.5in]{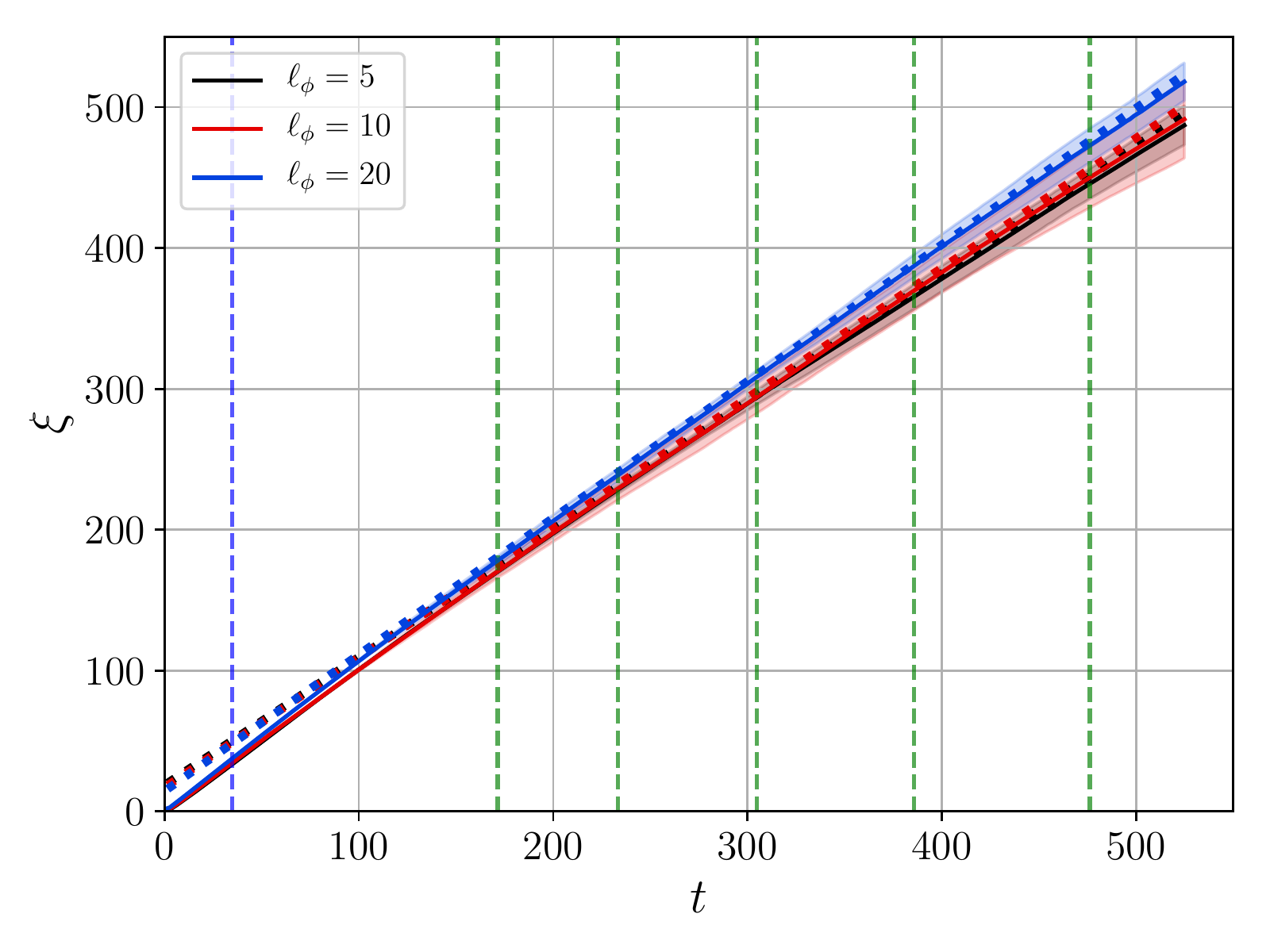}
    \caption{
    Mean string separation $\xi$ (defined in Eq.~(\ref{e:xisDef})) from 4k simulations with all initial field   
    correlation lengths $\ell_{\phi}$. 
    The solid line represents the mean over realizations of $\xi$ at each time, with 
    the shaded regions showing the 1-$\sigma$ variation.
    Also shown as dotted lines are the linear fits to the form of Eq.~(\ref{e:xiSca}), 
    whose parameters and uncertainties are shown in Table \ref{tab:slopes}. 
    The blue vertical dashed line is the end of the core growth period ($t_{\rm cg}$), after which strings maintain their physical width, 
    and the green ones are the boundaries of the fitting ranges.
    \label{fig:xiScaling}}
 \end{figure}

Fig.~\ref{fig:xiScaling} shows the results for the mean string separation $\xi$ for $4k$ simulations with different initial correlation lengths. Graphs of $\xi$ against $t$ for all runs are shown in the Supplemental Material.  
Consistent with our earlier  simulations \cite{Lopez-Eiguren:2017dmc}, 
after a relatively short period of relaxation, 
$\xi$ asymptotes to a line that can be well fitted with\footnote{Note that $\be$ as defined here is the slope of the
comoving string separation $\xi/a$ plotted against conformal time $\tau = 2 t/a$.} 
\ben
\label{e:xiSca}
\xi = 2\beta(t-\tOff). 
\een
This is the standard scaling model. 
The scaling value of the length density parameter is $\ze_0 = 1/4\be^2$.
 
We measure the parameters $\beta$ and $\tOff$ with a linear fit 
over four ranges in conformal time, defined by a vector of boundary times 
$\tau_{\rm b}=(6,7,8,9,10)\tau_s$, and $\tau_s\eta = [25, 50, 100]$ for 
$N = [1k,2k,4k]$.
We choose times in the last half of the conformal time range 
to minimise biases from the initial conditions.
The standard deviation of the central values of the parameters in the different fit ranges 
can be used to 
give an estimate of the combined uncertainty 
due to the approach to scaling and the lattice spacing: 
later fits will be closer to the scaling value, but more affected by the lattice spacing, 
which is equal to the inverse mass $(2\eta)^{-1}$ at the end of the simulation.
The standard deviation of the central values between different $\ell_\phi$ 
gives an estimate of the uncertainty    
due to the initial correlation length.
The two uncertainties are added in quadrature to give an estimate of the systematic error $\Delta\be_\text{sys}$, 
which is dominated by the uncertainty due to the variation in initial correlation lengths. 
The total uncertainty is obtained from adding the statistical and systematic uncertainties in quadrature. 
The means and uncertainties for the standard scaling parameters can be found in Tables~\ref{tab:slopes} and \ref{tab:slopes_errors}.

 \begin{table}
 \centering
 \resizebox{\columnwidth}{!}{%
\begin{tabular}{|c|c|c|c|c|} 
\hline
$\Delta t_{\rm fit}$ &  $\ell_{\phi}$& $t_0$  & $\beta$  & $\zeta_0$ \\
 \hline
171.42 - 233.33 &   5 & -8.94 $\pm$ 2.74 & 0.47 $\pm$ 0.01 & 1.12 $\pm$ 0.05 \\
171.42 - 233.33 &   10 & -7.48 $\pm$ 8.47 & 0.48 $\pm$ 0.03 & 1.11 $\pm$ 0.15 \\
171.42 - 233.33 &   20 & -10.00 $\pm$ 1.81 & 0.49 $\pm$ 0.01 & 1.04 $\pm$ 0.05 \\
\hline
233.33 - 304.76 &   5 & -16.46 $\pm$ 10.27 & 0.46 $\pm$ 0.02 & 1.20 $\pm$ 0.11 \\
233.33 - 304.76 &   10 & -19.64 $\pm$ 4.54 & 0.45 $\pm$ 0.02 & 1.22 $\pm$ 0.13 \\
233.33 - 304.76 &   20 & -12.59 $\pm$ 13.29 & 0.49 $\pm$ 0.02 & 1.06 $\pm$ 0.09 \\
\hline
304.76 - 385.71 &   5 & -29.83 $\pm$ 11.13 & 0.44 $\pm$ 0.02 & 1.30 $\pm$ 0.13 \\
304.76 - 385.71 &   10 & -12.32 $\pm$ 20.42 & 0.47 $\pm$ 0.03 & 1.16 $\pm$ 0.16 \\
304.76 - 385.71 &   20 & -12.93 $\pm$ 15.52 & 0.49 $\pm$ 0.03 & 1.07 $\pm$ 0.12 \\
\hline
385.71 - 476.19 &   5 & -27.32 $\pm$ 27.07 & 0.44 $\pm$ 0.03 & 1.28 $\pm$ 0.17 \\
385.71 - 476.19 &   10 & -34.48 $\pm$ 39.63 & 0.45 $\pm$ 0.05 & 1.31 $\pm$ 0.31 \\
385.71 - 476.19 &   20 & -23.79 $\pm$ 16.37 & 0.47 $\pm$ 0.03 & 1.12 $\pm$ 0.13 \\
\hline
\end{tabular} %
}
  \caption{\label{tab:slopes} 
Numerical values of the fit parameters for the 4k runs fitted over the conformal time ranges given after Eq. (\ref{e:xiSca}), 
  shown in physical time as $\Delta t_{\rm fit}$.
The fit parameters $t_0$ and $\beta$ pertain to Eq.~(\ref{e:xiSca}), with $\ze_0 = 1/4\be^2$. 
The values are computed averaging over the 4 different realisations and the 
computation of the uncertainties is described after Eq.~(\ref{e:xiSca}).}

\end{table}

 \begin{table}
 \centering
\resizebox{\columnwidth}{!}{%
\begin{tabular}{|c|c|c|c|c|c|c|} 
\hline
$N$ & $\beta \pm \Delta\beta$ & $\Delta\beta_{\rm stat}$ & $\Delta\beta_{\rm sys}$& $\zeta_0 \pm \Delta\zeta_0$ & $\Delta\zeta_{0,\rm stat}$ & $\Delta\zeta_{0,\rm sys}$  \\
 \hline
$1k$ & 0.499 $\pm$ 0.042 & 0.031 & 0.028 & 1.02 $\pm$ 0.17 & 0.13 & 0.11  \\
$2k$ & 0.486 $\pm$ 0.036  & 0.030 & 0.019 & 1.07 $\pm$ 0.16 & 0.13 & 0.08 \\
$4k$ & 0.467 $\pm$ 0.037  & 0.030 & 0.021 & 1.17 $\pm$ 0.20 & 0.17 & 0.11  \\
\hline
\end{tabular} %
}
  \caption{\label{tab:slopes_errors} 
Central values and estimated uncertainties of the standard scaling parameters $\be$ and $\ze_0=1/4\be^2$ for all box sizes. 
The decomposition into statistical and systematic uncertainties, as discussed in the text after Eq.~(\ref{e:xiSca}), is also shown.}

\end{table}

We now turn to the alternative models recently put forward: logarithmic \cite{Gorghetto:2018myk,Vaquero:2018tib,Kawasaki:2018bzv,Buschmann:2019icd}, and inverse logarithmic \cite{Martins:2018dqg} correction to scaling
\ben
\zeta(t)=\zeta^*_0+\alpha^* \log(\eta t), \quad \zeta(t)=\zeta_0' + {\alpha'}/{ \log(\eta t)},
\label{eq:logcorr}
\een
 where $\zeta^*_0$, $\zeta_0'$, $\alpha^*$ and $\alpha'$ are the fitting parameters.
 We performed fits over the four ranges used previously, 
 using the same method to estimate uncertainties.  
The mean values and uncertainties for the parameters can be found in Table~\ref{tab:alternates}. 

The uncertainties include zero, and are 
apparently inconsistent with reports of a logarithmic correction with coefficient 
$\al^* \simeq 0.2$ \cite{Gorghetto:2018myk,Kawasaki:2018bzv}.
It is interesting to examine why. If the strings are scaling in the sense that the mean string separation $\xi$ is increasing linearly, 
the string density parameter $\ze$ behaves as  
\ben
\ze = \frac{t^2}{4\be^2(t - \tOff)^2} \simeq \ze_0\left( 1 + 2\frac{\tOff}{t} \right). 
\label{e:ZetCur}
\een
The uncorrected estimator 
approaches its asymptotic value slowly,  
resembling the behaviour of a logarithm with a positive coefficient,  
\ben
\alpha^*(t_f) = -2\zeta(\tOff/t_f)(1 - \tOff/t_f)^{-1},
\label{e:AlpStaPre}
\een
where $t_f$ is a time at which the fit is carried out. 
We find that taking $t_f$ to be the final time in the fit range gives the best fit. 
If $\tOff < 0$, the approach is from lower (``underdense'') values of $\zeta$, giving positive values of $\alpha^*$, and vice versa.
Hence an apparent logarithmic growth parameter $\alpha^* \simeq 0.2$ \cite{Gorghetto:2018myk,Kawasaki:2018bzv} 
is produced for runs where the initial string configurations are biased towards $t_0/t_f \simeq - 0.1$.  
Our initial conditions cover both positive and negative values of $\alpha^*$, 
and are consistent with $\alpha^* = 0$ as $\tOff/t_f \to 0$. 
The parameter $\alpha'$ similarly takes both signs and is 
consistent with zero as $\tOff/t_f \to 0$. 
The constant terms in the alternative models are consistent with standard scaling 
$\zeta^*_0, \zeta'_0 \simeq 1$ as $\tOff/t_f \to 0$. 
The standard scaling parameter $\beta$ depends only weakly on $\tOff/t_f$.  This 
effect is included in our uncertainty, and is smaller than the statistical fluctuations.
More information is given in the Supplemental Material.

\begin{figure}[htbp]
    \centering
    \includegraphics[width=3.35in]{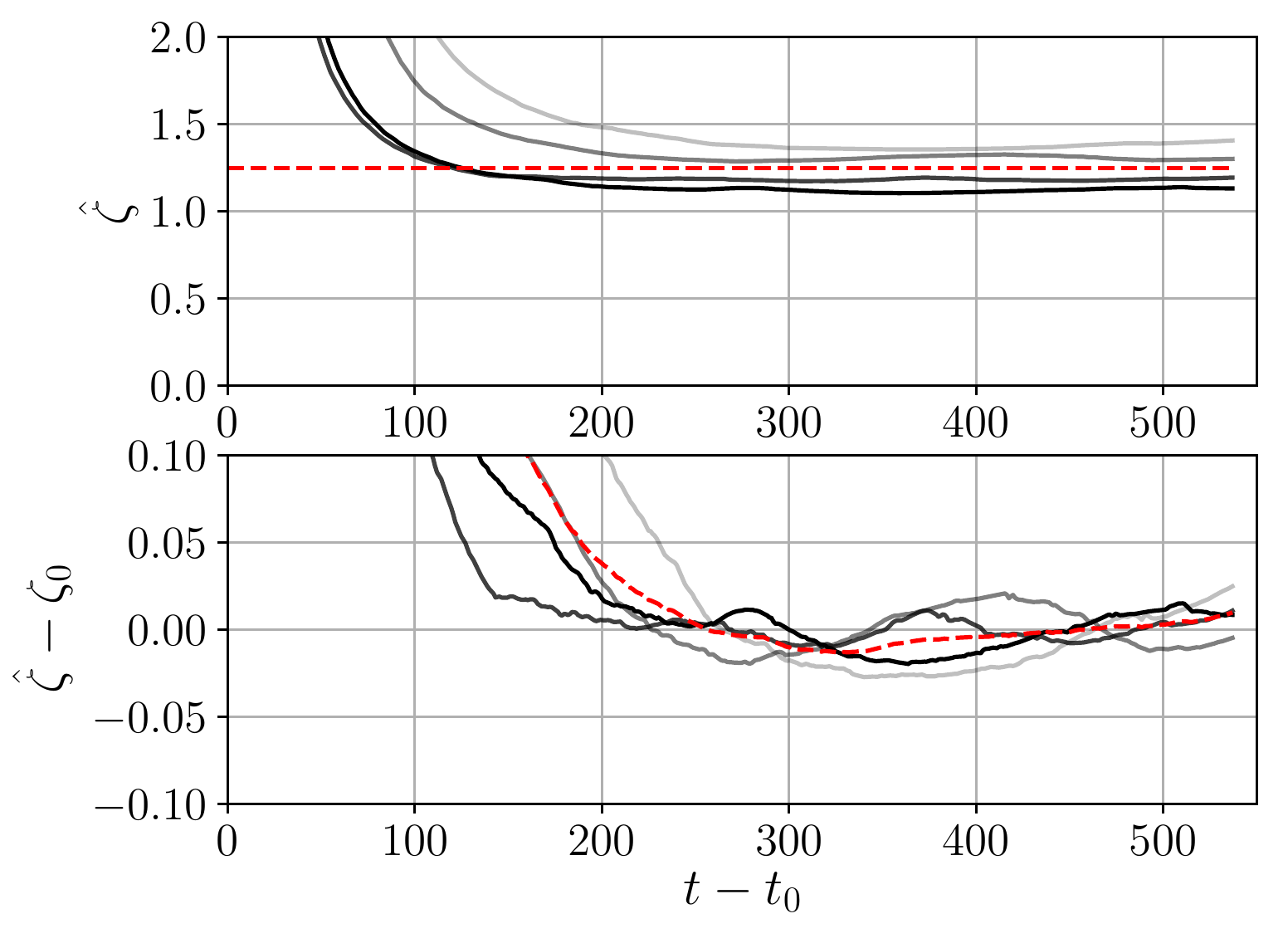}
     \caption{
      Top: string length density parameter $\hat\zeta$ (see Eq.~(\ref{e:xiSca})) plotted against offset time $t - t_0$, for all $4k$ runs with initial correlation length $\ell_\phi = 5$. The dashed line shows the mean of $\zeta_0 = 1/4\beta^2$. 
      Bottom: residuals ($\hat\zeta-\zeta_0$) plotted against offset time $t - t_0$,  for the same simulations. Individual runs are showed in solid lines and the residuals between the mean $\zeta_0$ and $\hat\zeta$ in dashed.    \label{fig:zeta_k}}
 \end{figure}

We also explore the possibility of a small drift away from standard scaling in the residuals, by using
the length density parameter estimator 
\ben
\zeEst = {\ell_s (t-\tOff)^2}/{\mathcal{V}}={(t -\tOff)^2}/{\xi^2}\,,
\label{zeta2}
\een
where $\tOff$ is the best fit value from the fit (\ref{e:xiSca}) for $\xi(t)$. 
In Fig.~\ref{fig:zeta_k}  we plot $\zeEst$ against $t-\tOff$ for the 4 runs with $\ell_{\phi}\eta=5$. 
The figure gives a clear impression of $\zeEst$ 
tending to an asymptotically constant value.  The residuals to the standard scaling fit for $\ell_\phi\eta = 5$ are also shown in Fig.~\ref{fig:zeta_k}, with the mean shown as a dashed line.
We fit the residual to a constant plus a logarithm according to 
\ben
\zeEst(t) - \zeta_0 = \zeta_\text{r} + \alpha_\text{r} \log( \eta t ),
\label{eq:alpha}
\een
where $\alpha_\text{r}$ and $\zeta_\text{r}$ are fitting parameters,  
fitted over the four ranges in conformal time described earlier.

 \begin{table}
 \centering
 \resizebox{\columnwidth}{!}{%
\begin{tabular}{|c|c|c|c|c|c|c|} 
\hline
$N$ &   $\zeta^*_0$  & $\alpha^*$ &   $\zeta_0'$  & $\alpha'$ &  $\zeta_r\ (\times 10^{-2})$  & $\alpha_r\ (\times 10^{-2})$ \\
 \hline
$1k$ &  1.7 $\pm$ 1.0 & -0.14 $\pm$ 0.21 & 0.55$\pm$ 0.69 & 2.4$\pm$ 3.3 &    0.0 $\pm$ 1.3 & -0.02 $\pm$ 0.31 \\
$2k$ &    0.88 $\pm$ 0.60 & 0.03 $\pm$ 0.11 &   1.18 $\pm$ 0.58 & -0.8 $\pm$ 3.0 &    0.2 $\pm$ 1.6 & -0.04 $\pm$ 0.33\\
$4k$ &    0.42 $\pm$ 0.59 & 0.11 $\pm$ 0.11 &    1.66 $\pm$ 0.68 & -3.6 $\pm$ 3.8 &    0.2 $\pm$ 1.5 & -0.03 $\pm$ 0.26 \\
\hline
\end{tabular} %
}
  \caption{\label{tab:alternates} 
  Numerical values of the fit parameters of the logarithmic correction, inverse logarithmic correction, and residuals as presented in Eqs.~(\ref{eq:logcorr}) and (\ref{eq:alpha}) respectively. Fitting ranges and error estimates were obtained following the same prescription as in the previous tables. 
}

\end{table}

The measured values of $\zeta_\text{r}$ and  $\alpha_\text{r}$  are given in Table~\ref{tab:alternates}, along with the uncertainties. 
They are consistent with zero, 
and give a tight bound on any logarithmic growth in the 
string length density parameter in the residual.

Having determined that standard scaling is the best model, we explore
the uncertainty due to the finite lattice volume.
We average the fit parameters over initial correlation lengths and fit ranges at each 
lattice size, and then perform a linear extrapolation in $1/L\eta$. 
Our final result for the length density parameter is\footnote{ We have also performed simulation with constant comoving width, and observe a similar behaviour, with $\zeta_0 = 1.34 \pm 0.22$. See the Supplemental Material.}
\ben
\label{e:zetaFinal}
\ze_0 = \zzMean \pm \zzErr.
\een
The coefficient of any logarithm in the residuals is 
\ben
\label{e:alphaFinal}
\al_{\rm r} = (\LogCorMean \pm \LogCorErr) \times 10^{-2},
\een
consistent with zero. 
The dominant error is statistical.


\mysection{Conclusions}
\label{sec:Con}
In this paper we have  investigated the scaling density of axion strings, 
prompted by recent claims of a logarithmic increase in the string length density parameter $\ze$  
\cite{Fleury:2015aca,Klaer:2017qhr,Klaer:2017ond,Gorghetto:2018myk,Vaquero:2018tib,Kawasaki:2018bzv,Buschmann:2019icd}.

We have fitted the string length density from our simulations with three two-parameter models:
the standard scaling model with  the usual  time offset $\tOff$ to account for the initial string evolution and an asymptotically constant length density parameter $\zeta_0$;
a model with a logarithmically increasing $\ze$;
and a model with an inverse logarithmic correction.  
By linear fits to the mean string separation $\xi$, 
we obtain a well-determined result for the parameter $\ze_0$, given in Eq.~(\ref{e:zetaFinal}).
The coefficients of the logarithm and inverse logarithm 
can be understood in terms of the dependence of 
$\ze$ on the initial string evolution parameter $\tOff$, and describe a disguised approach to 
scaling for non-zero $\tOff$. 
We find they are consistent with zero when $\tOff/t_f\to 0$, where $t_f$ is the final fitting time. 
The constant terms in the models consistent with the standard scaling values.
A search for a logarithmic correction to the residuals of the standard scaling model 
gives a tight upper bound on the magnitude of its coefficient (\ref{e:alphaFinal}):
our $3\sigma$ limit on a logarithmic correction to the string density parameter 
is $|\alpha| < 0.94\times10^{-2}$.

We conclude that axion strings scale very well in the standard sense, and that between a 
$10^{12}$ GeV PQ phase transition and the QCD transition at $100$ MeV, 
any logarithmic correction to the string density parameter $\ze_0 \simeq 1$ must be less than about 0.5.

An implication of the confirmation of standard scaling, important for network modelling \cite{Martins:2018dqg}, 
is that the energy loss rate per unit length of string must increase at the same rate as the effective string tension.

The tight constraint on the logarithmic correction also has implications for attempts to extend the dynamic range of 
global string simulations 
\cite{Fleury:2015aca,Klaer:2017qhr,Klaer:2017ond} by using frustrated strings \cite{Hill:1987bw}. 
Frustrated string models have fields with both global and local symmetries, and the string resembles a global string with an 
Abelian Higgs string at the core.  
The effect is to decouple the string tension $\mu$ and the axion decay constant $\fa$, so that $\ka = \mu/\pi\fa^2$ can be chosen to be greater than 1. 
As the effective tension of an axion string is  $\mu_\text{a} \simeq \pi \fa^2 \ln(\xi \eta)$,
it was argued that a simulation with frustrated strings would effectively reach a string separation 
$\xi \sim \eta^{-1} \exp(\ka)$. 

It was found 
that there was an increase in the length density parameter $\ze$ with 
the ratio $\ka = \mu/\pi f_\text{a}^2$, apparently saturating at $\ze \simeq 20$ around $\ka \simeq 50$ \cite{Klaer:2017ond}. 
This is far above our O(1) upper bound on $\zeta$ at the QCD scale,  
 casting doubt on the effectiveness of frustrated strings as a generic model of 
axion strings at large separations.
Hence one should not extrapolate 
the axion number density $n_\text{ax}$ to $\kappa \sim 70$. 
From Fig.~6 (right) of Ref.~\cite{Klaer:2017ond}
one can estimate that ${n_\text{ax}}/{n_\text{mis}} \simeq 0.5$ at $\ka = 1$, 
where $n_\text{mis}$ is the angle-averaged 
number density produced by the misalignment mechanism 
\cite{Abbott:1982af,Dine:1982ah,Preskill:1982cy,Bae:2008ue,Wantz:2009it,Borsanyi:2016ksw,Klaer:2017ond}. 
This is consistent with the directly-measured values 
reported by other groups \cite{Kawasaki:2014sqa,Vaquero:2018tib}.
This value is about 60\% 
of the extrapolated value \cite{Klaer:2017ond}, 
suggesting that the value of the axion dark matter mass of about $25\;\mu\text{eV}$ \cite{Klaer:2017ond} should be revised 
upwards by about 50\% in
scenarios based on PQ symmetry-breaking by a gauge singlet. 
We leave a more precise estimate for future work. 

Finally, we note that frustrated string models \cite{Fleury:2015aca,Klaer:2017qhr,Klaer:2017ond} may be viable if 
the PQ symmetry-breaking is accompanied by the breaking of a U(1) gauge symmetry.
The difference in the axion dark matter mass estimates 
between the models 
implies that the detection of an axion and 
an accurate measurement of its mass 
could distinguish between them.

\begin{acknowledgments}
We are grateful for fruitful discussions with M.~Kawasaki, J.~Redondo, K.~Saikawa, T.~Sekiguchi, G.~Villadoro, M.~Yamaguchi and J.~Yokoyama. 
MH (ORCID ID 0000-0002-9307-437X) acknowledges support from the Science and Technology Facilities Council (grant number ST/L000504/1). JL (ORCID ID 0000-0002-1198-3191) and JU (ORCID ID 0000-0002-4221-2859) acknowledge support from Eusko Jaurlaritza (IT-979-16) and PGC2018-094626-B-C21 (MCIU/AEl/FEDER,UE). ALE (ORCID ID 0000-0002-1696-3579) is supported by the Academy of Finland grant 286769. 
ALE is grateful to the Early Universe Cosmology group of the University of the Basque Country for their generous hospitality and useful discussions.  
This work has been possible thanks to the computational resources on the STFC DiRAC HPC facility obtained under the dp116 project. Our simulations also made use of facilities at the i2Basque academic network and CSC Finland.
\end{acknowledgments}

\bibliography{axion} 

\newpage
\hspace{10cm}

\FloatBarrier
\newpage
\section{Supplemental Material}
\setcounter{figure}{0}
\renewcommand{\thefigure}{S\arabic{figure}}

\subsection{Infinite volume extrapolation of fit parameters}
\begin{figure}[htbp]
    \centering
    \includegraphics[width=3.5in]{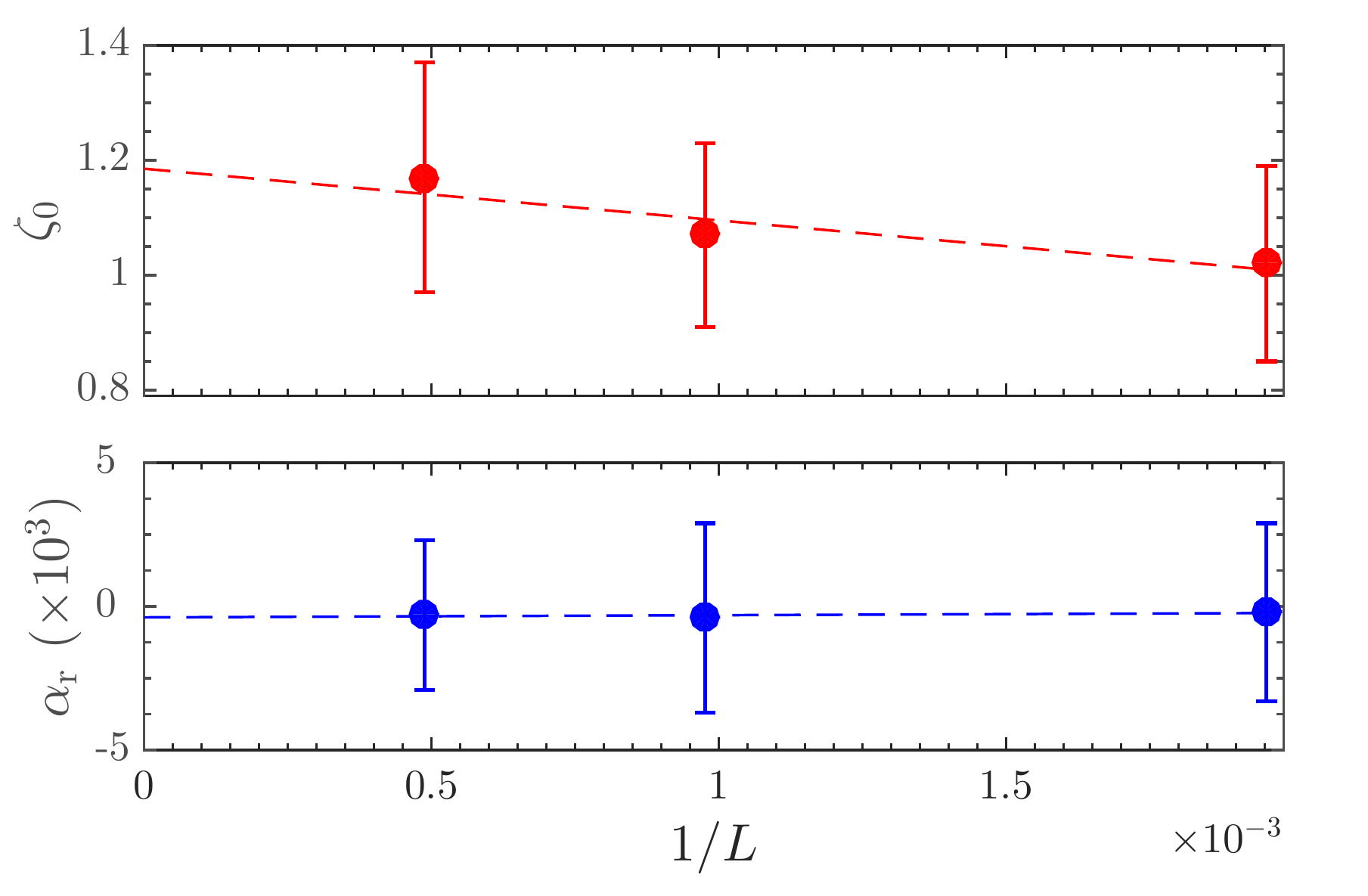}
     \caption{Central values and estimated uncertainties of the fit parameters $\ze_0$ and $\al_{\rm r}$ (see Tables~\ref{tab:slopes_errors} and \ref{tab:alternates}) plotted against $1/L$, where $L = N\delta x$ is the comoving side-length of the simulation box.
          The dashed lines are linear fits to the data points.
     \label{fig:extrapolation}}
 \end{figure}

In Fig.~\ref{fig:extrapolation} we show the central values of the fit parameters $\ze_0$ and $\al_{\rm r}$ (see Eq.~\ref{eq:alpha}) as well as the 1-$\sigma$ uncertainties for $1k$, $2k$ and $4k$ simulations against $1/L$. The values can be seen in Tables~\ref{tab:slopes_errors} and \ref{tab:alternates}. As explained in the main text we perform a linear extrapolation to obtain the final results for $\ze_0$ and $\al_{\rm r}$, these linear extrapolations are shown in the plots as dashed lines.

\subsection{String separation and density from all simulations}

In Fig.~\ref{f:SupMatGrid} we show plots of 
$\xi$ against $t$ and $\zeta$ against $\log(t\eta)$ for all simulations, 
where $\xi$ is  the mean string separation  (\ref{e:xisDef}), 
$\zeta$ is the string length density parameter (\ref{eq:zeta}), 
$t$ is cosmic time, and $\eta$ the expectation value of the field. 
We additionally show two sets of $4k$ simulations with $k=1008$, $\tau_{\rm ini}\eta=70$, $\tau_{\rm diff}\eta=90$, $\tau_{\rm cg}\eta=307.4$, and $\tau_{\rm ini}\eta=90$, $\tau_{\rm diff}\eta=110$, $\tau_{\rm cg}\eta=339.85$.  These simulations were performed on a different architecture from the others, necessitating a slight reduction 
in the simulation volume. They are designed to confirm that $\zeta$ also tends to its scaling value from above in $4k$ simulations.

\subsection{Dependence of fit parameters on initial conditions}

In Fig.~\ref{f:SupMatAlpha} we show the dependence of the fit parameters $\beta$, 
$\alpha^*$ and $\zeta_0^*$, and  $\alpha^{\prime}$ and $\zeta_0^{\prime}$ (\ref{eq:logcorr}) on the fit parameter $t_0/t_f$, where $\tOff$ is the time offset in the fit  (\ref{e:xiSca})
and $t_f$ is the end of the fitting period. 
The dotted line in the middle figure is Eq.~(\ref{e:AlpStaPre}), with $t_f$ taken as the end of the fitting period (the rightmost dashed lines in Fig.~\ref{f:SupMatGrid}). The solid black line represents our final result for the length density parameter (\ref{e:zetaFinal}) with the 1-$\sigma$ variations represented as shaded regions.

\subsection{Constant comoving width simulations}
We also include the corresponding results for simulations with constant comoving width, i.e., using the Press-Ryden-Spergel method \cite{Press:1989yh}: Table~\ref{tab:SupMatslopes_errors}  is analogous to Table~\ref{tab:slopes_errors} in the main text,  i.e., it shows the central values and estimated uncertainties (as discussed in the main text after Eq.~(\ref{e:xiSca})) for $\beta$ and $\zeta_0=1/4\beta^2$, but for simulations with constant comoving width.  Also, the figures corresponding to Figs~\ref{f:SupMatGrid} and \ref{f:SupMatAlpha} are  Figs.~\ref{f:SupMatGrids0} and \ref{f:SupMatAlphas0}, respectively.  The combined value for the length density parameter $\zeta_0$ for the constant comoving width case is:
\begin{equation}
\zeta_0=1.34\pm0.22
\label{zeta0s0}
\end{equation}

\begin{table}[h]
 \centering
\begin{tabular}{|c|c|c|c|c|c|c|} 
\hline
$N$ & $\beta \pm \Delta\beta$ & $\Delta\beta_{\rm stat}$ & $\Delta\beta_{\rm sys}$& $\zeta_0 \pm \Delta\zeta_0$ & $\Delta\zeta_{0,\rm stat}$ & $\Delta\zeta_{0,\rm sys}$  \\
 \hline
$1k$ & 0.459 $\pm$ 0.040 & 0.040 & 0.027 & 1.21 $\pm$ 0.22 & 0.17 & 0.14  \\
$2k$ & 0.447 $\pm$ 0.037  & 0.031 & 0.020 & 1.27 $\pm$ 0.21 & 0.18 & 0.12 \\
$4k$ & 0.440 $\pm$ 0.035  & 0.029 & 0.020 & 1.31 $\pm$ 0.21 & 0.17 & 0.12  \\
\hline
\end{tabular} 
  \caption{\label{tab:SupMatslopes_errors} 
Central values and estimated uncertainties of the standard scaling parameters $\be$ and $\ze_0=1/4\be^2$ for all box sizes for simulations with comoving string width. 
The decomposition into statistical and systematic uncertainties, as discussed in the main text after Eq.~(\ref{e:xiSca}), is also shown.}
\end{table}

\subsection{Summary}

\begin{figure}[t!]
  \centering
 \includegraphics[width=0.32\textwidth,clip]{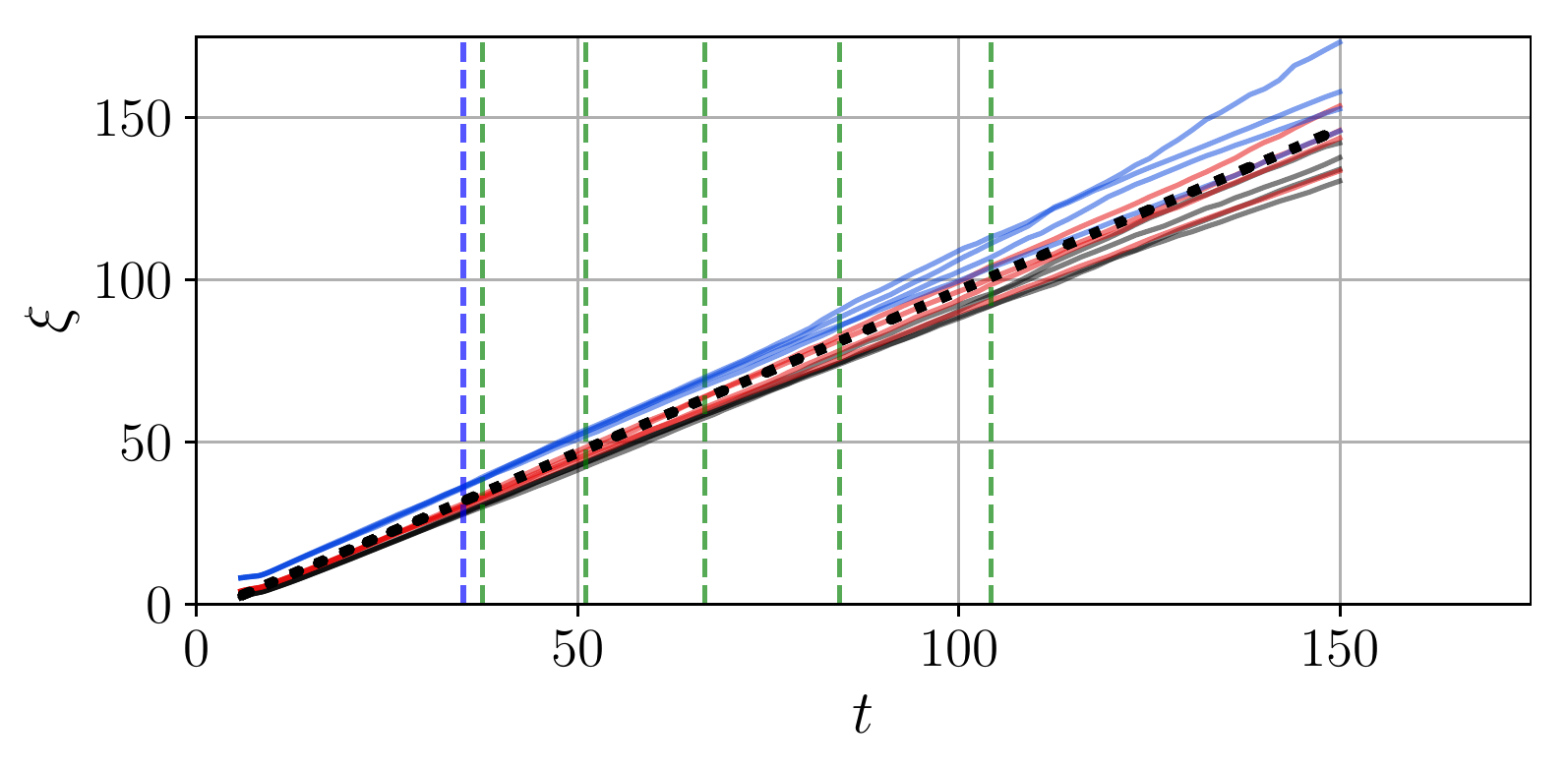}\hfill
\includegraphics[width=0.32\textwidth,clip]{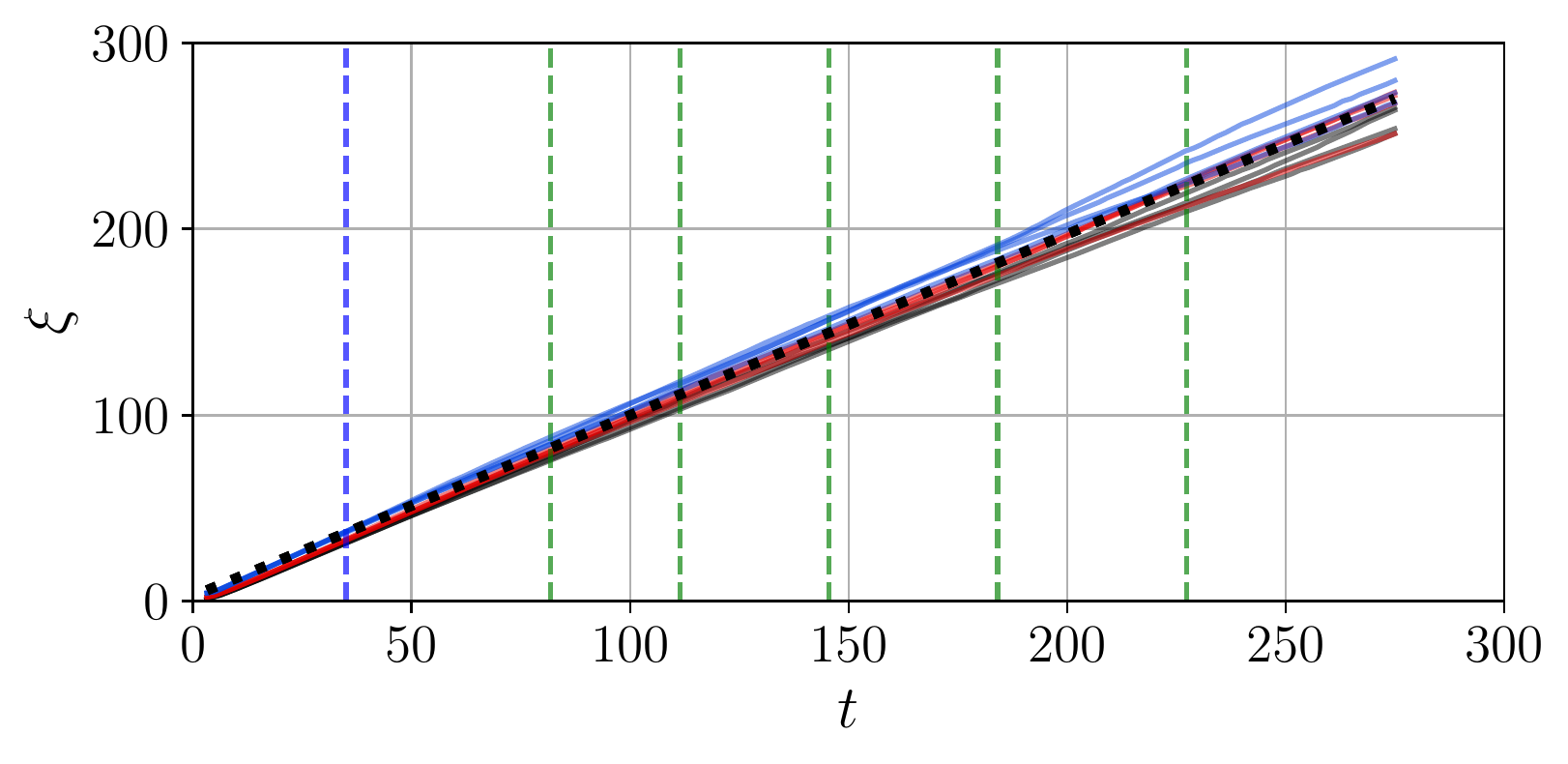}\hfill
\includegraphics[width=0.32\textwidth,clip]{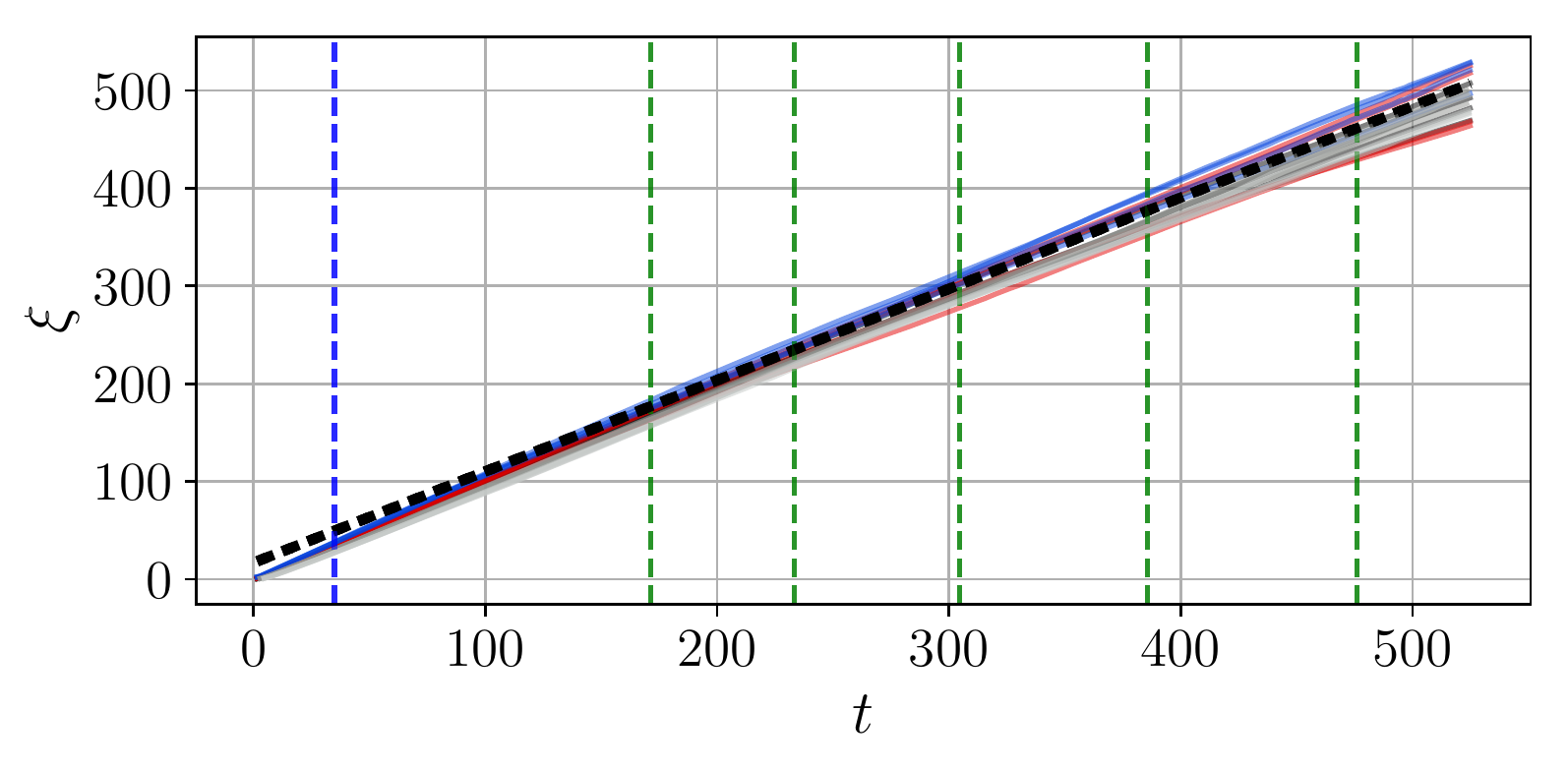}\\\hfill

\includegraphics[width=0.32\textwidth,clip]{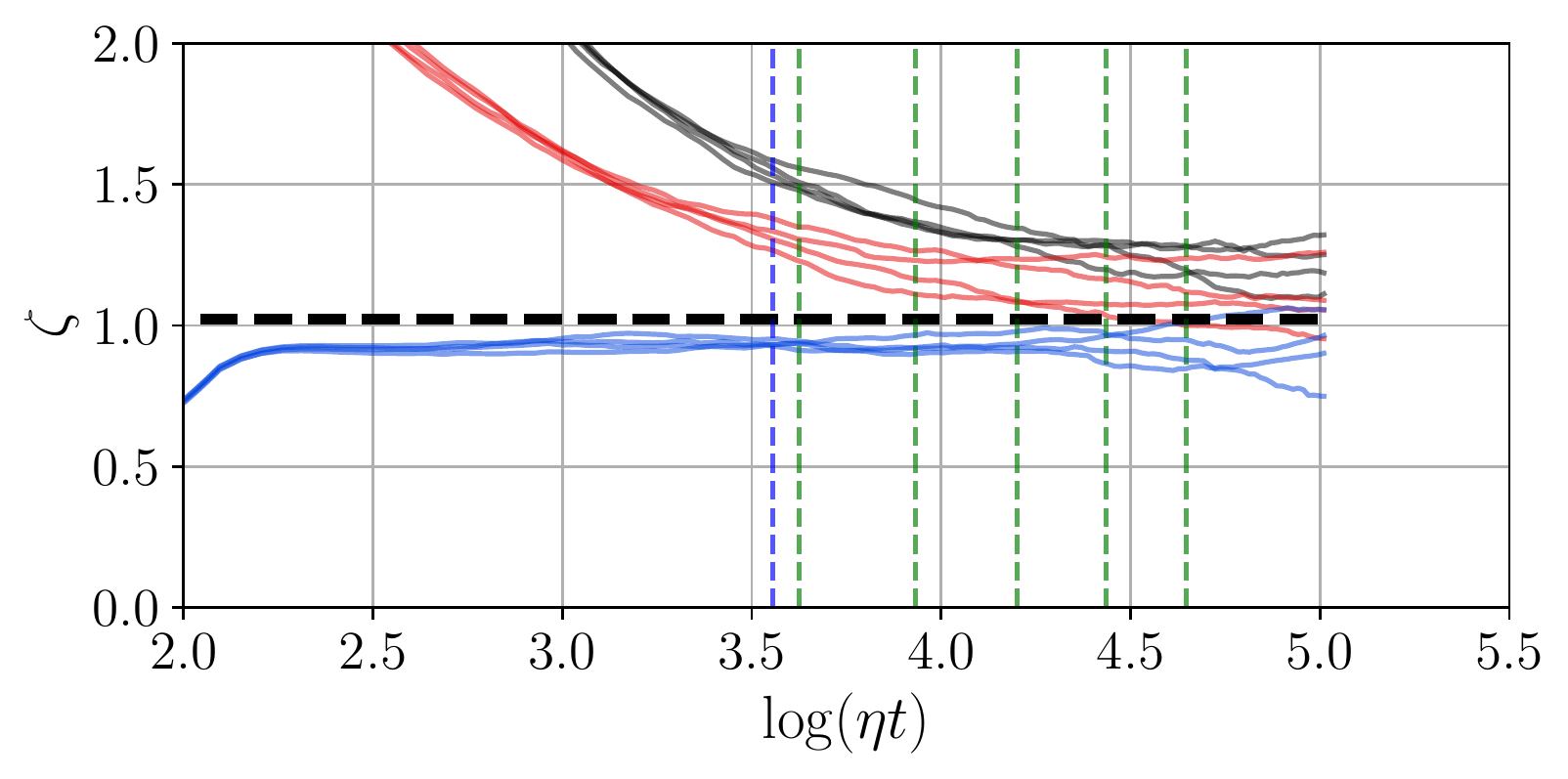}\hfill
\includegraphics[width=0.32\textwidth,clip]{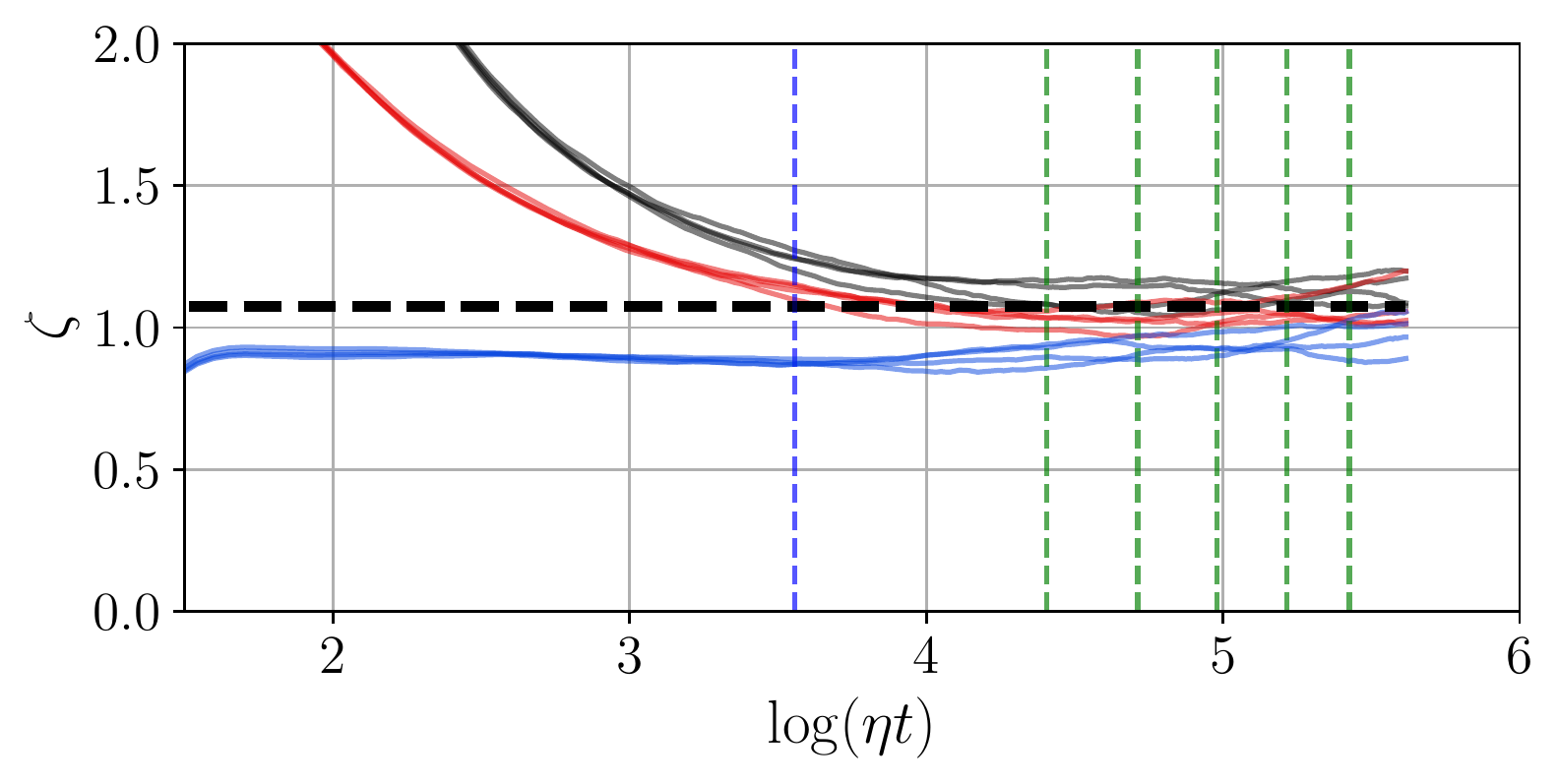}\hfill
\includegraphics[width=0.32\textwidth,clip]{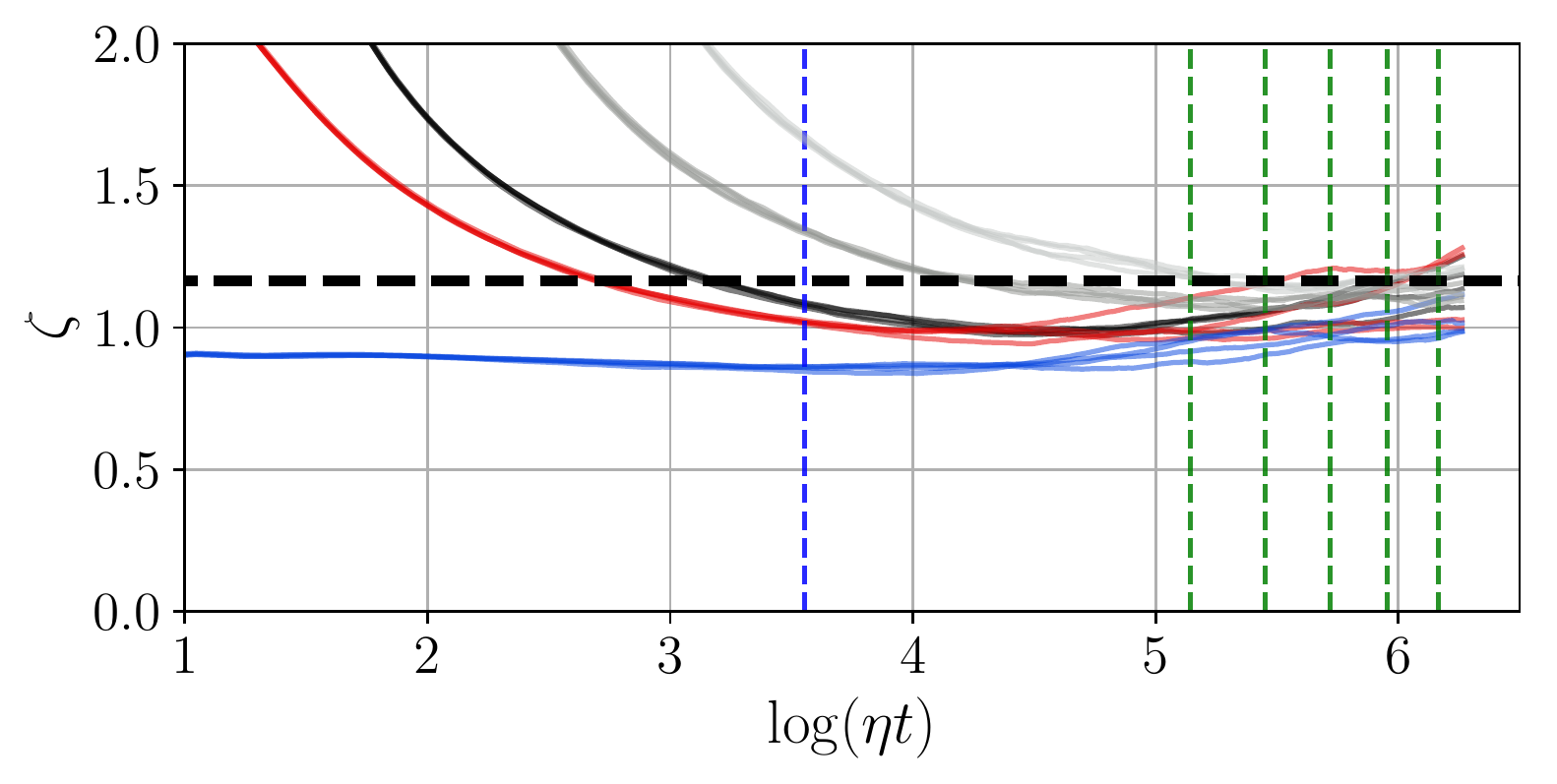}\\\hfill

\includegraphics[width=0.32\textwidth,clip]{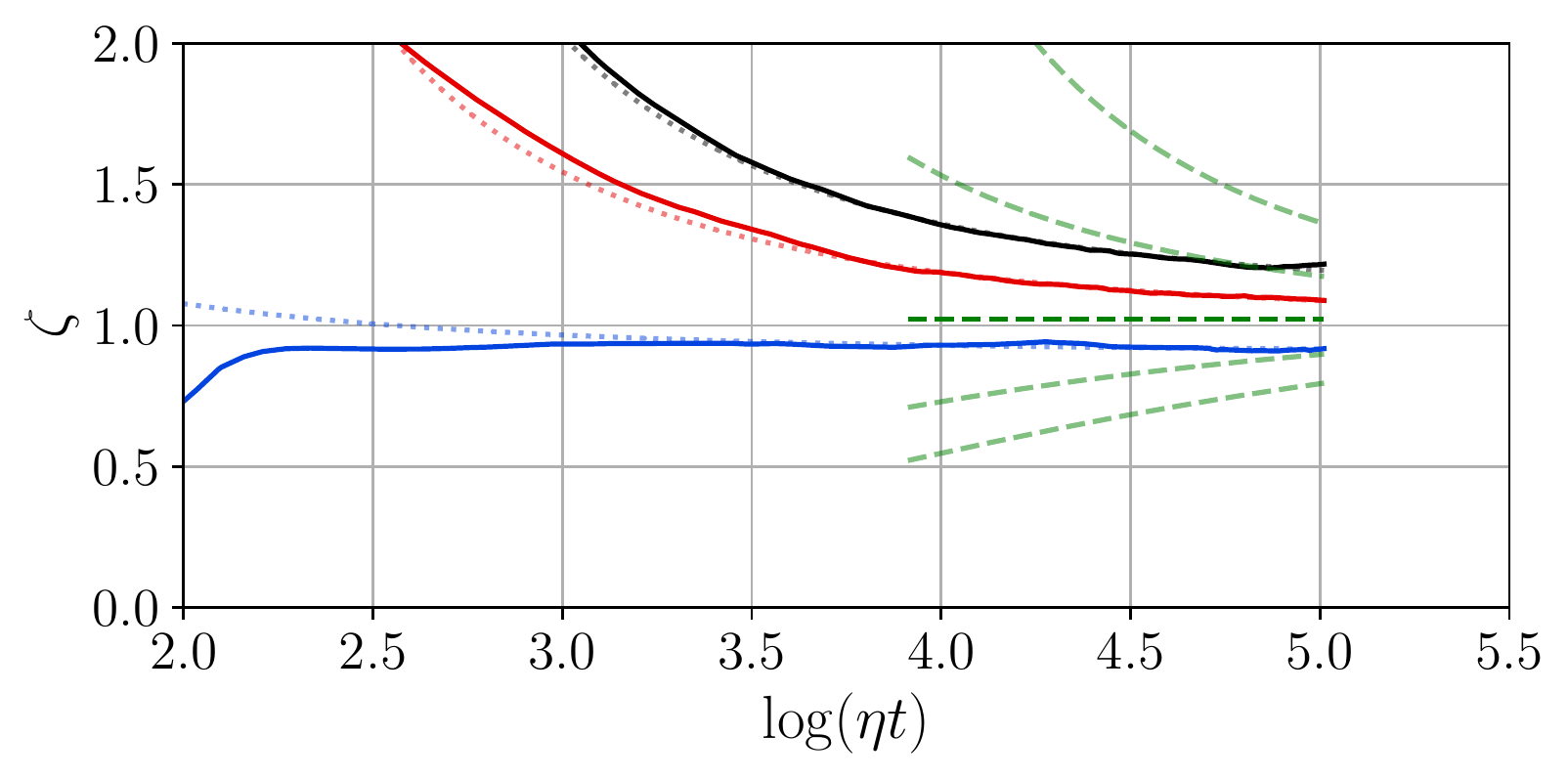}\hfill
\includegraphics[width=0.32\textwidth,clip]{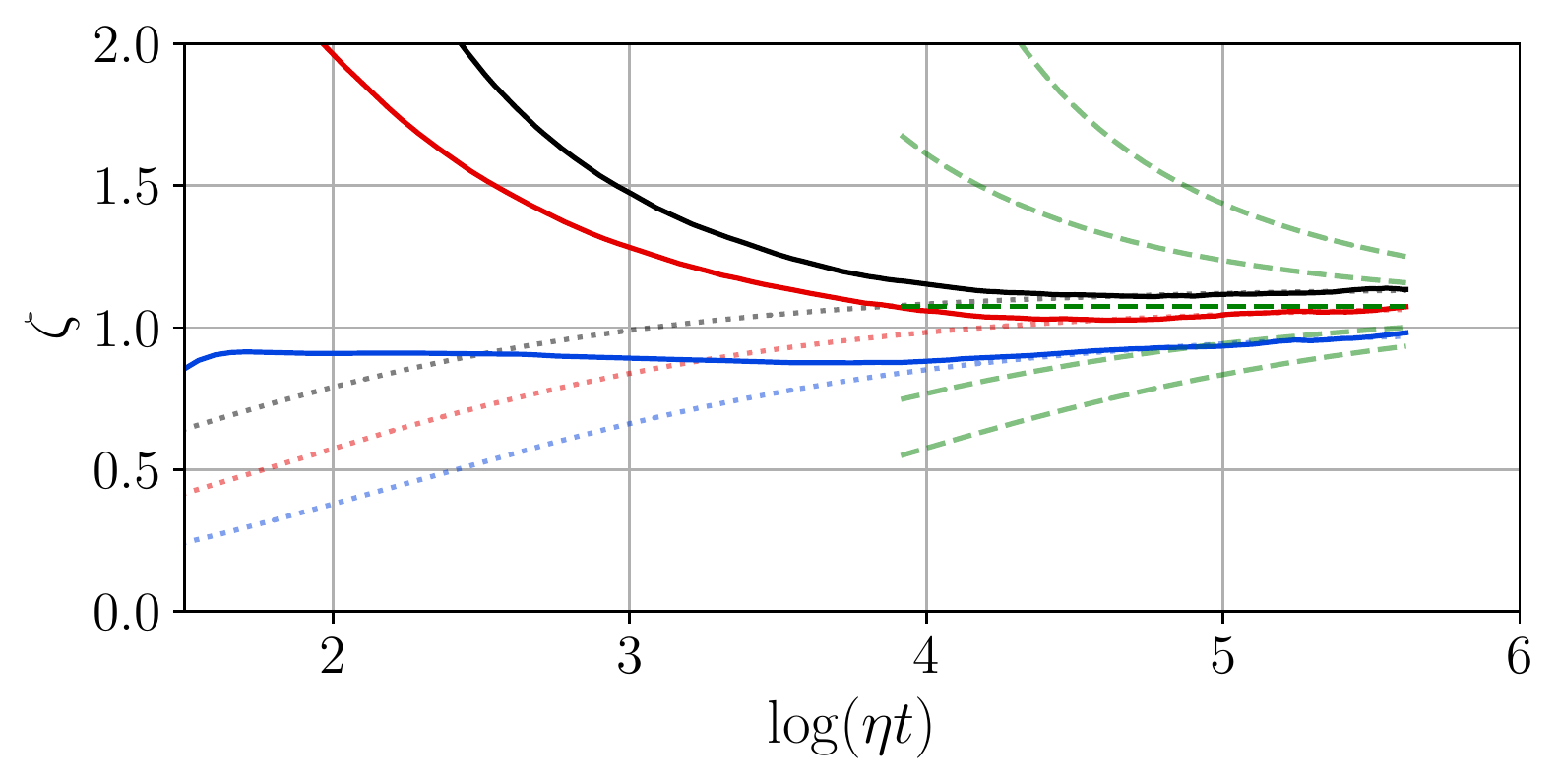}\hfill
\includegraphics[width=0.32\textwidth,clip]{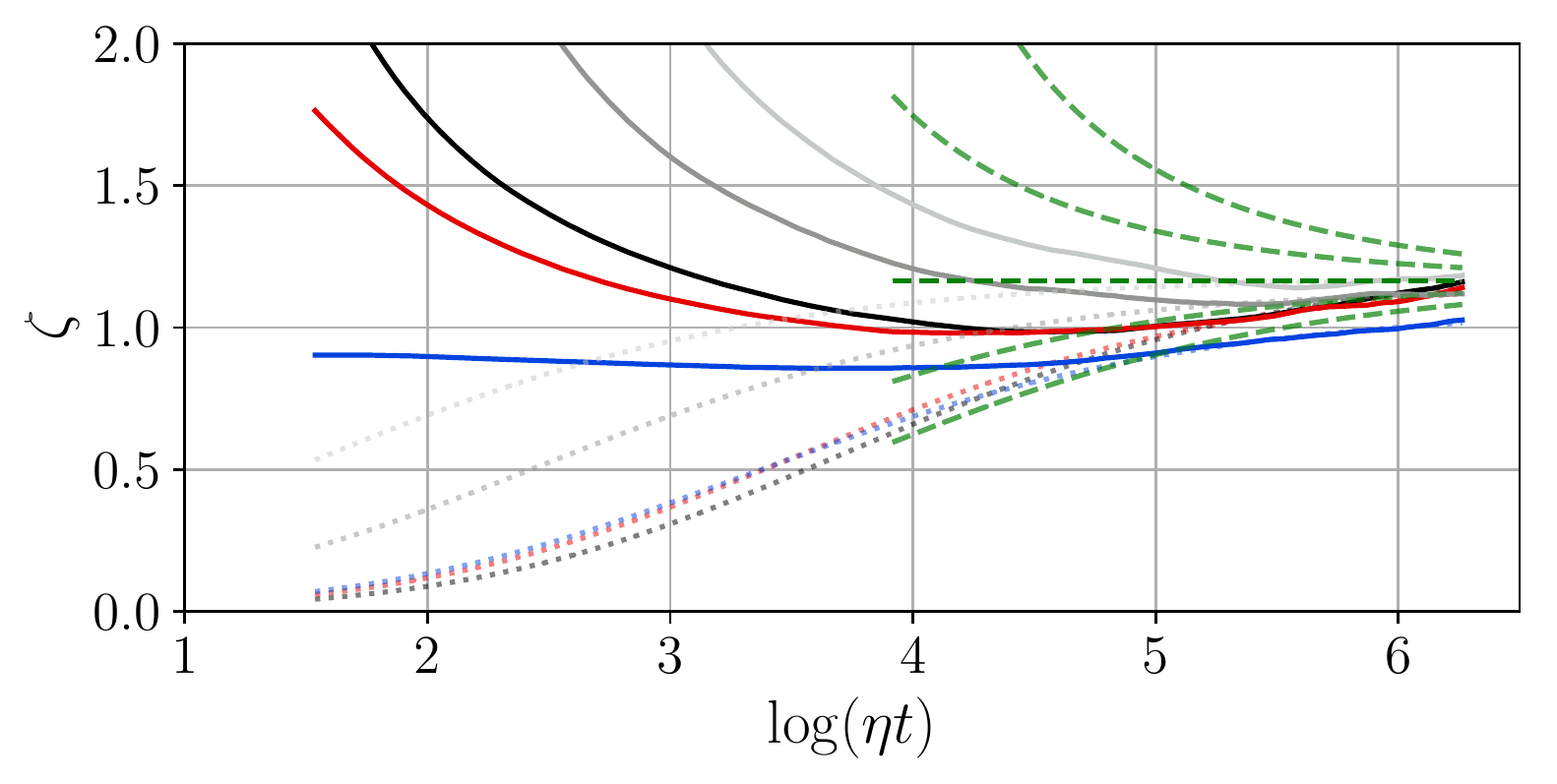}\\\hfill
\caption{
We show data from the $1k$, $2k$ and $4k$ simulations from left to the right, where the black lines represent simulations with initial field correlation length $l_{\phi}=5$, red ones with $l_{\phi}=10$ and blue ones with $l_{\phi}=20$. The $4k$ panel also contains an additional set of simulations with $l_{\phi}=5$ and shifted initialisation times (see text for explanation) in grey. The blue vertical dashed lines are the end of the core growth period ($t_{\rm cg}$), after which strings maintain their physical width, and green ones are the boundaries of the fitting ranges.
Top row:  Mean string separation $\xi$ (see Eq.~\ref{e:xisDef}) for each run simulated. The black dashed line represents $\xi=2\beta(t-t_0)$, where $\beta$ and $t_0$ are computed as stated in the main text for each lattice size. The values of $\beta$ can be found in Table~\ref{tab:slopes_errors}. 
Middle row: The length density parameter $\zeta$ (see Eq.~(\ref{eq:zeta})) for each run. The black dashed lines represent $\zeta_0=1/4\beta^2$, where we used the values of $\beta$ shown in Table~\ref{tab:slopes_errors}. 
Bottom row: The length density parameter $\zeta$ (see Eq.~\ref{eq:zeta}) averaged over the four runs with the same initial field correlation length $l_{\phi}$. The dotted lines show the curves $\zeta=\zeta_0(1-t_0/t)^2$, with the values of $\zeta_0=1/4\beta^2$ and $t_0$ used in figures at the top row. The green dashed lines are the fit curves $\zeta=\zeta_0(1-t_0/t)^2$ with $\eta |t_0|=0,10,20$.
}
\label{f:SupMatGrid}
\end{figure}

\begin{figure}[b!]
  \centering
 \includegraphics[width=0.32\textwidth,clip]{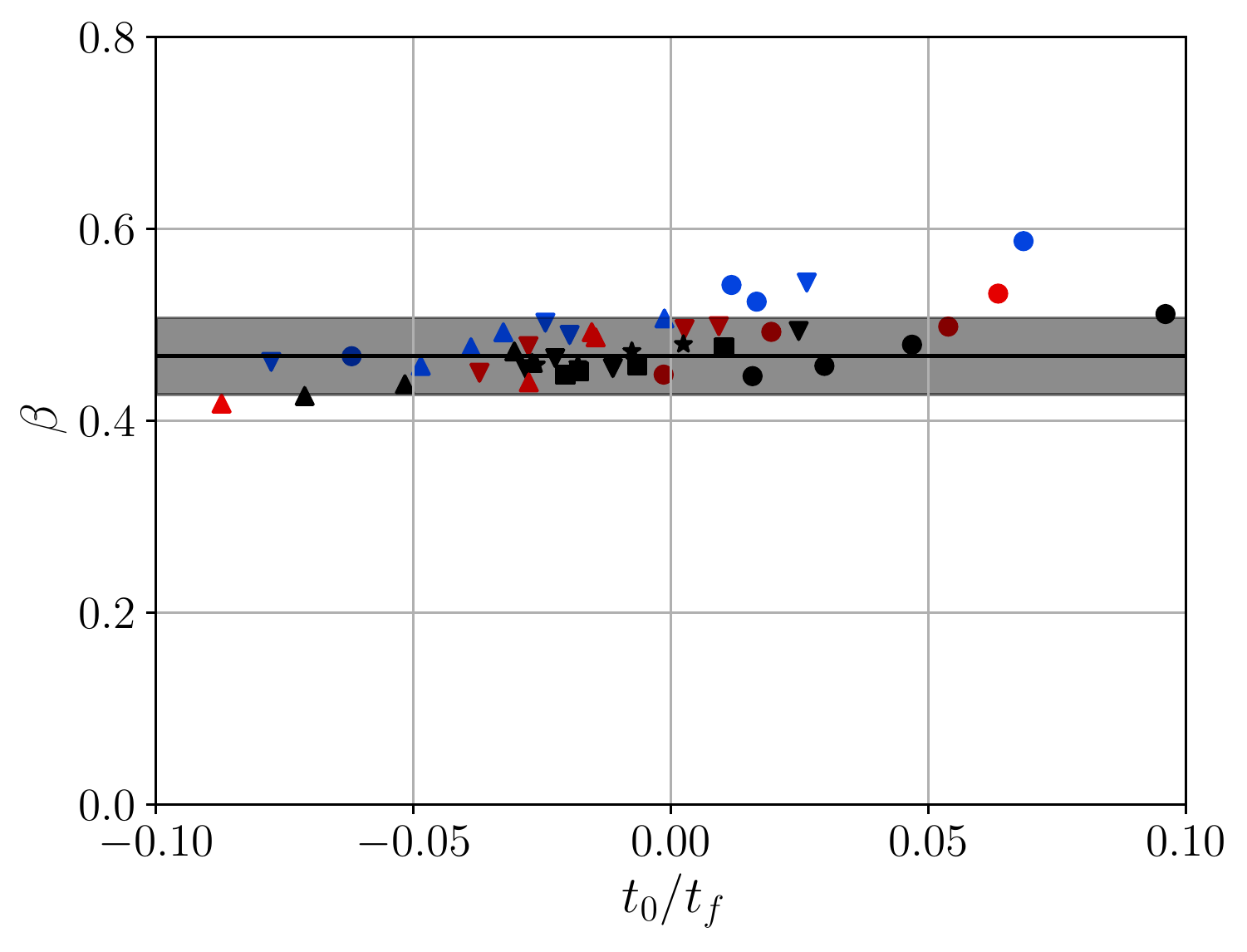}\hfill
\includegraphics[width=0.32\textwidth,clip]{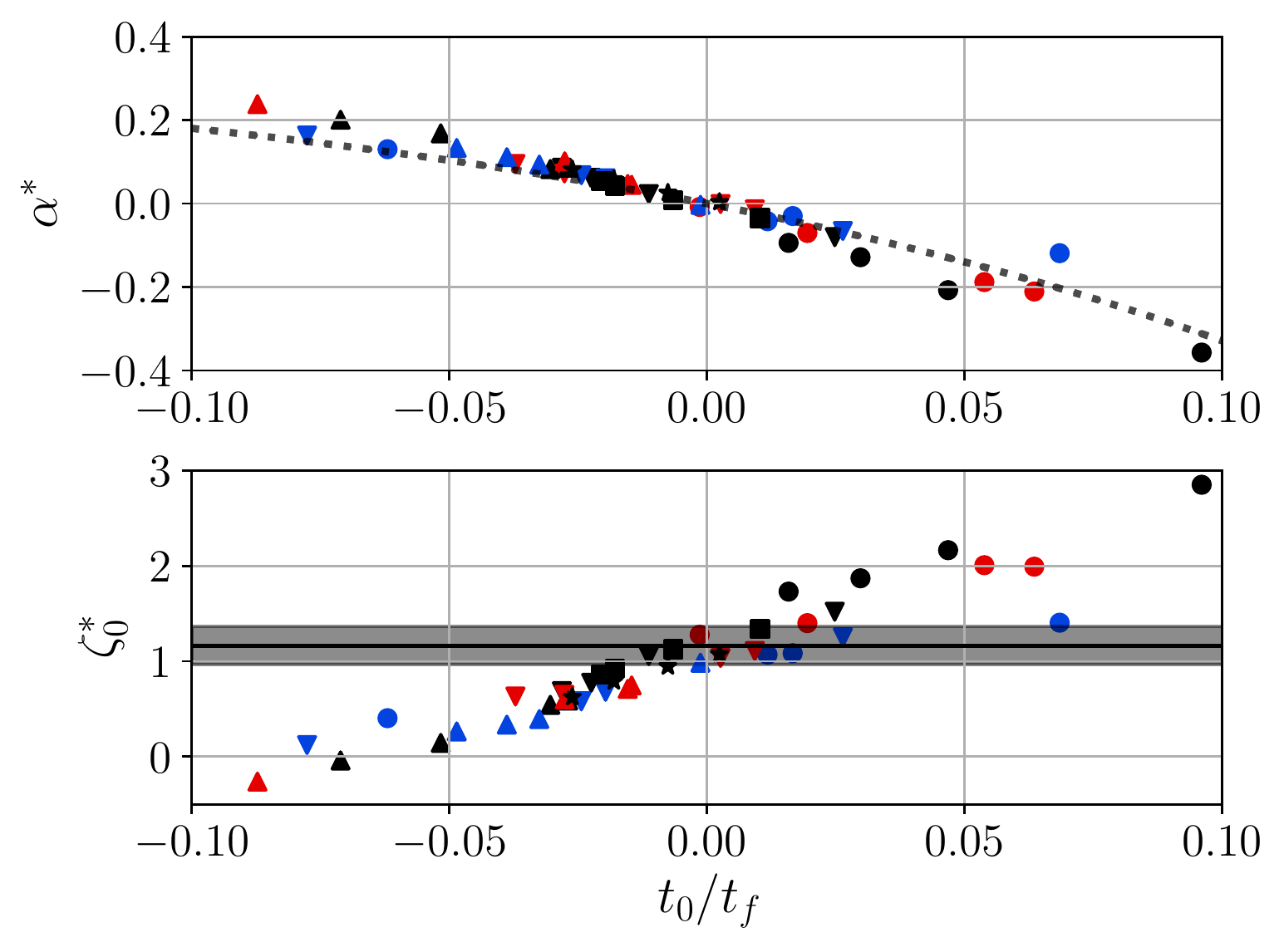}\hfill
\includegraphics[width=0.32\textwidth,clip]{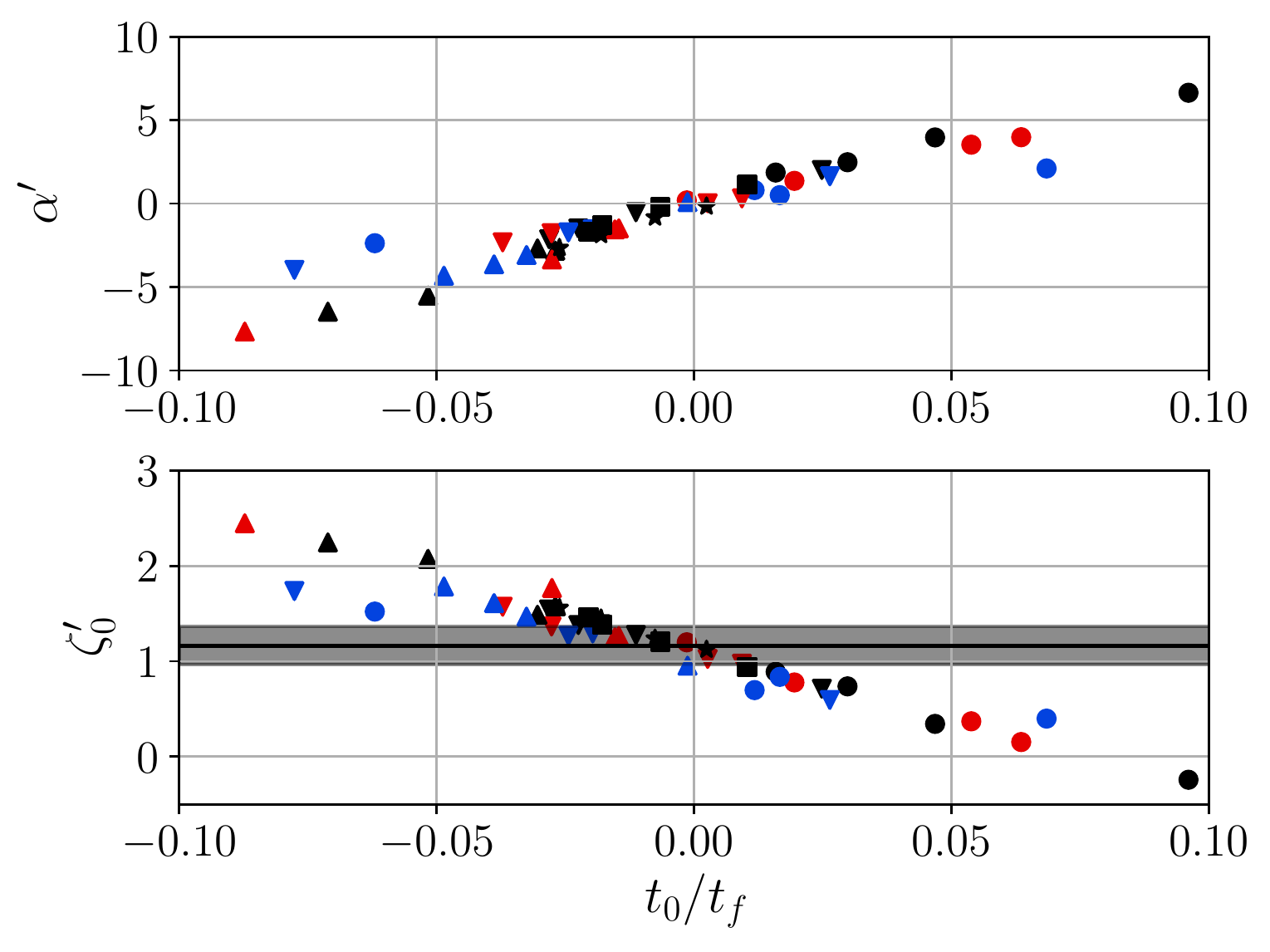}\\\hfill
\caption{Scatter plots showing the values of the fit parameters $\beta$ (see Eq.~\ref{e:xiSca}) on the left, $\alpha^*$ and $\zeta_0^*$ in the middle, and $\alpha^{\prime}$ and $\zeta_0^{\prime}$ (see Eq.~(\ref{eq:logcorr})) on the right, against $\tOff/t_f$, where $\tOff$ is the time offset in the fit  (\ref{e:xiSca}), and  
$t_f$ has been taken as the end of the fitting period (the rightmost dashed lines in Fig.~\ref{f:SupMatGrid}). The black line with the shaded region in the leftmost figure represents $\beta=1/2\sqrt{\zeta_0}$ with its 1-$\sigma$ variation, where the value has been taken to be the one in Eq.~(\ref{e:zetaFinal}) (i.e., $\zeta_0=1.19\pm 0.20$).  The black lines with the shaded region in the bottom of the other two figures are the corresponding $\zeta_0$ with its 1-$\sigma$ variation. The dotted line in the middle figure represents $\alpha^*(t_f)=-2\zeta(t_0/t_f)(1-t_0/t_f)^{-1}$.  The color code of the points represent the initial field correlation length where $l_{\phi}=5$ is black, $l_{\phi}=10$ is red and $l_{\phi}=20$ is blue. $1k$ simulations are represented as circles, $2k$ simulations as triangles pointing downwards, $4k$ simulations as triangles pointing upwards and $4k$ simulations with shifted initialisation times as stars ($\tau_{\rm ini}\eta=70$) and squares ($\tau_{\rm ini}\eta=90$)}
\label{f:SupMatAlpha}
\end{figure}

\begin{figure}[t!]
  \centering
 \includegraphics[width=0.32\textwidth,clip]{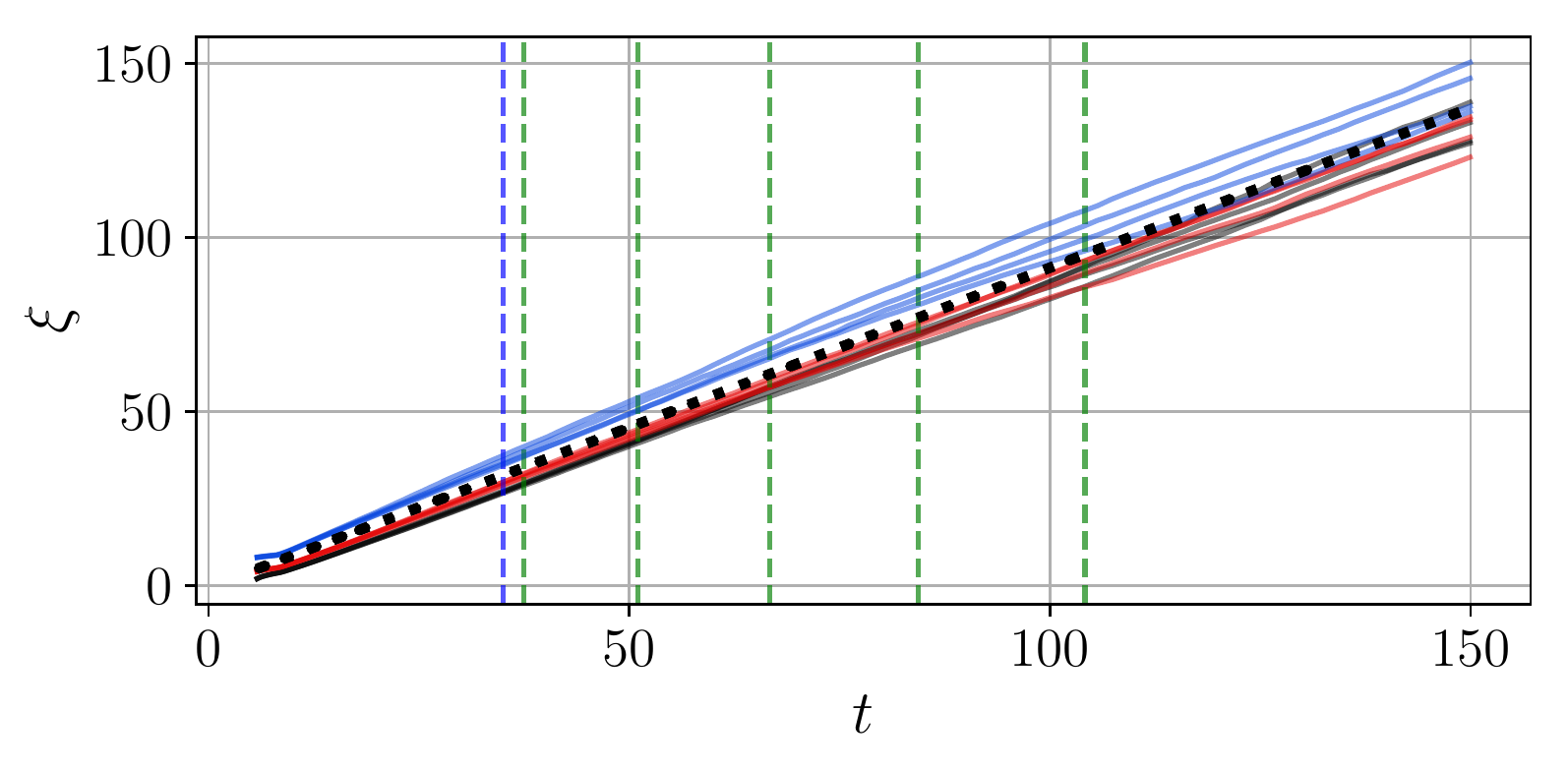}\hfill
\includegraphics[width=0.32\textwidth,clip]{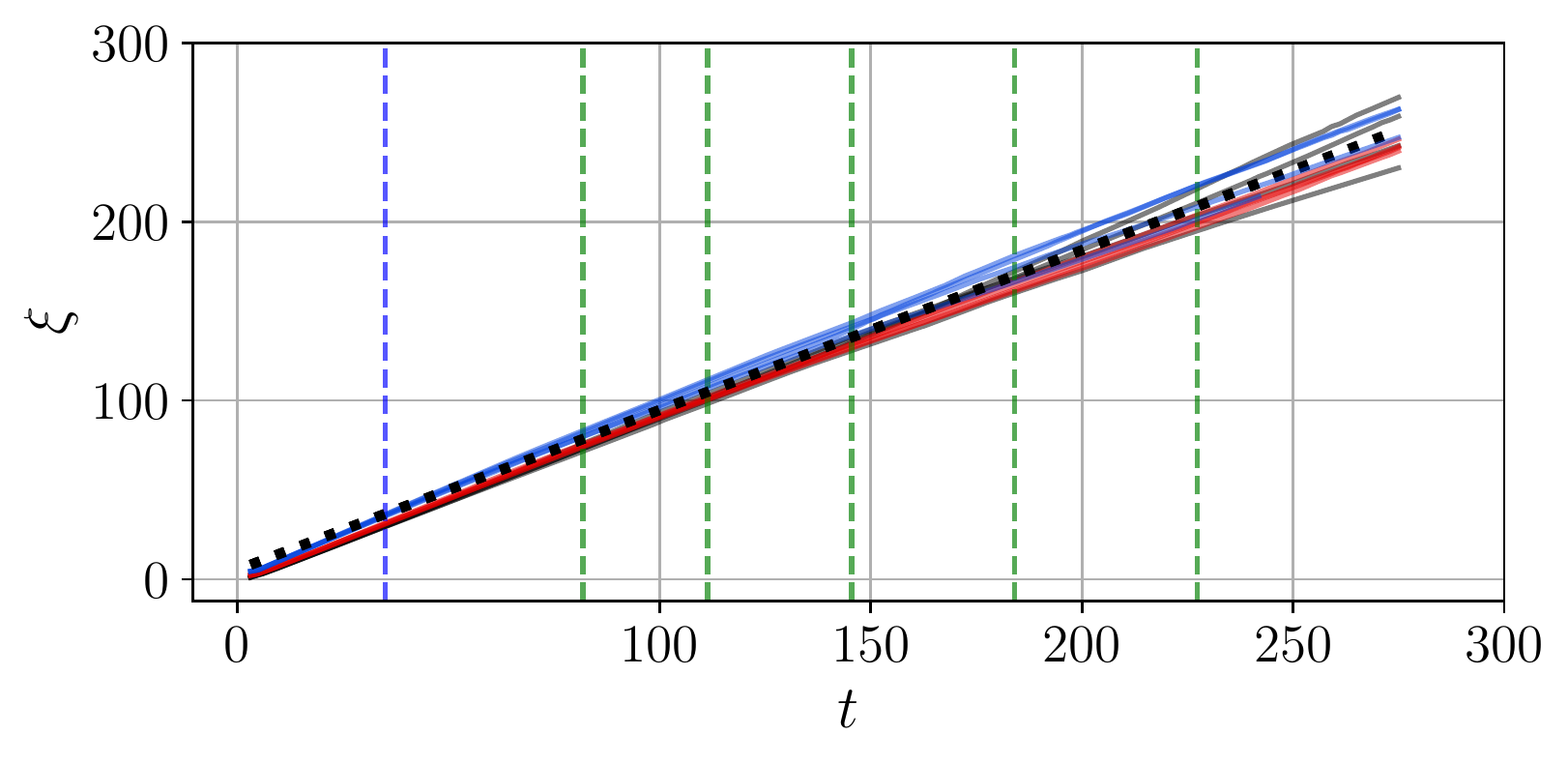}\hfill
\includegraphics[width=0.32\textwidth,clip]{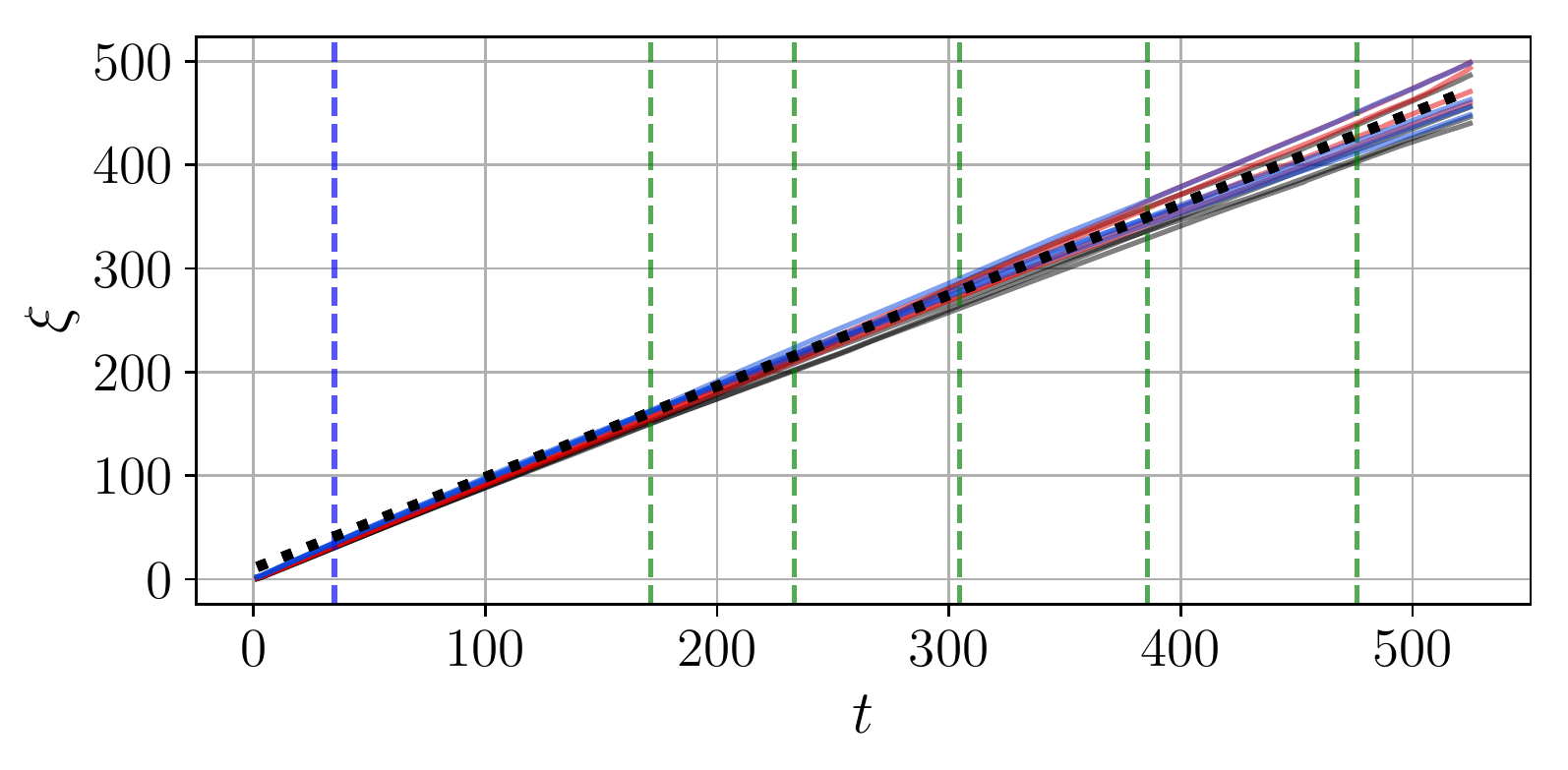}\\\hfill

\includegraphics[width=0.32\textwidth,clip]{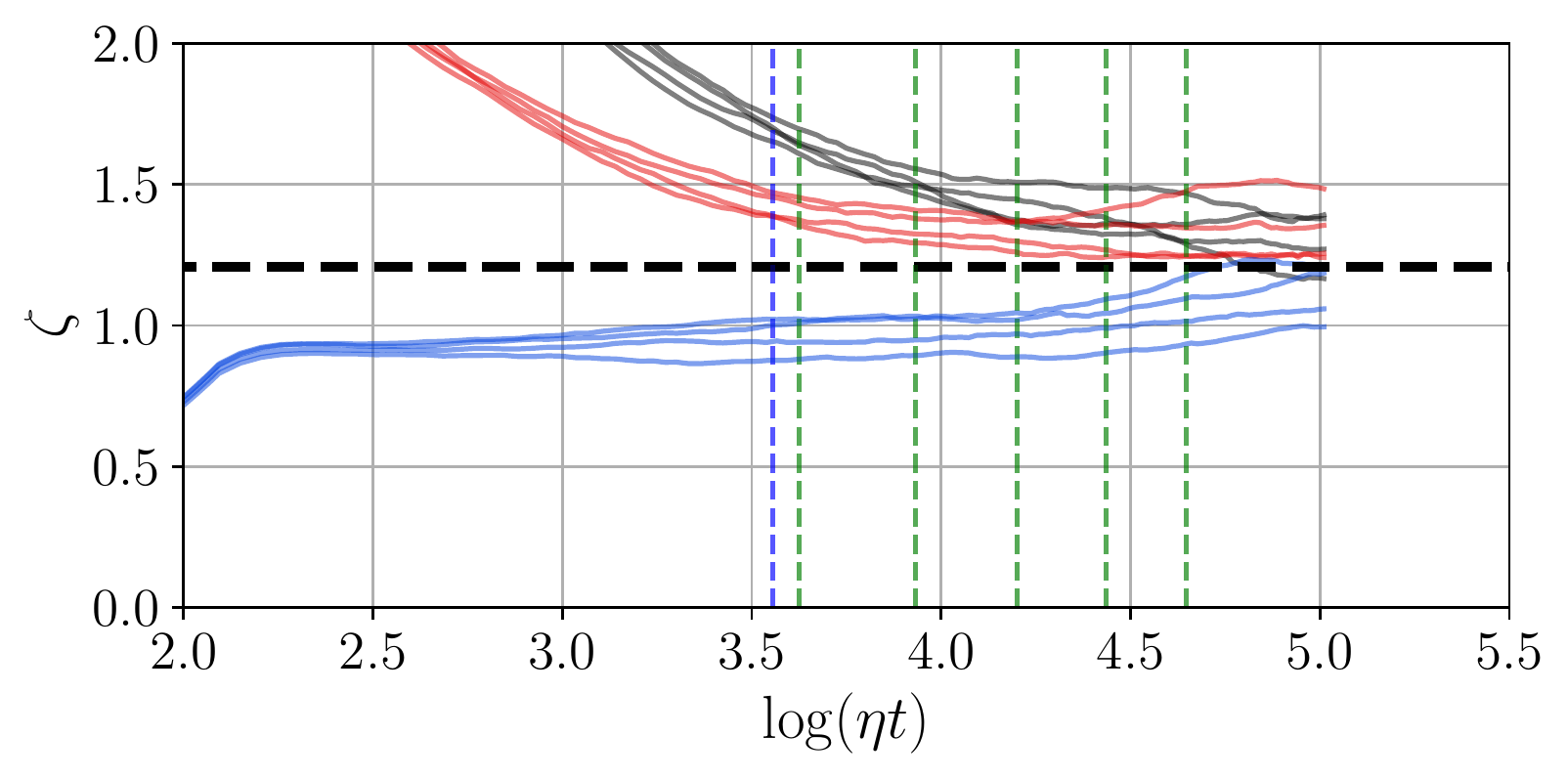}\hfill
\includegraphics[width=0.32\textwidth,clip]{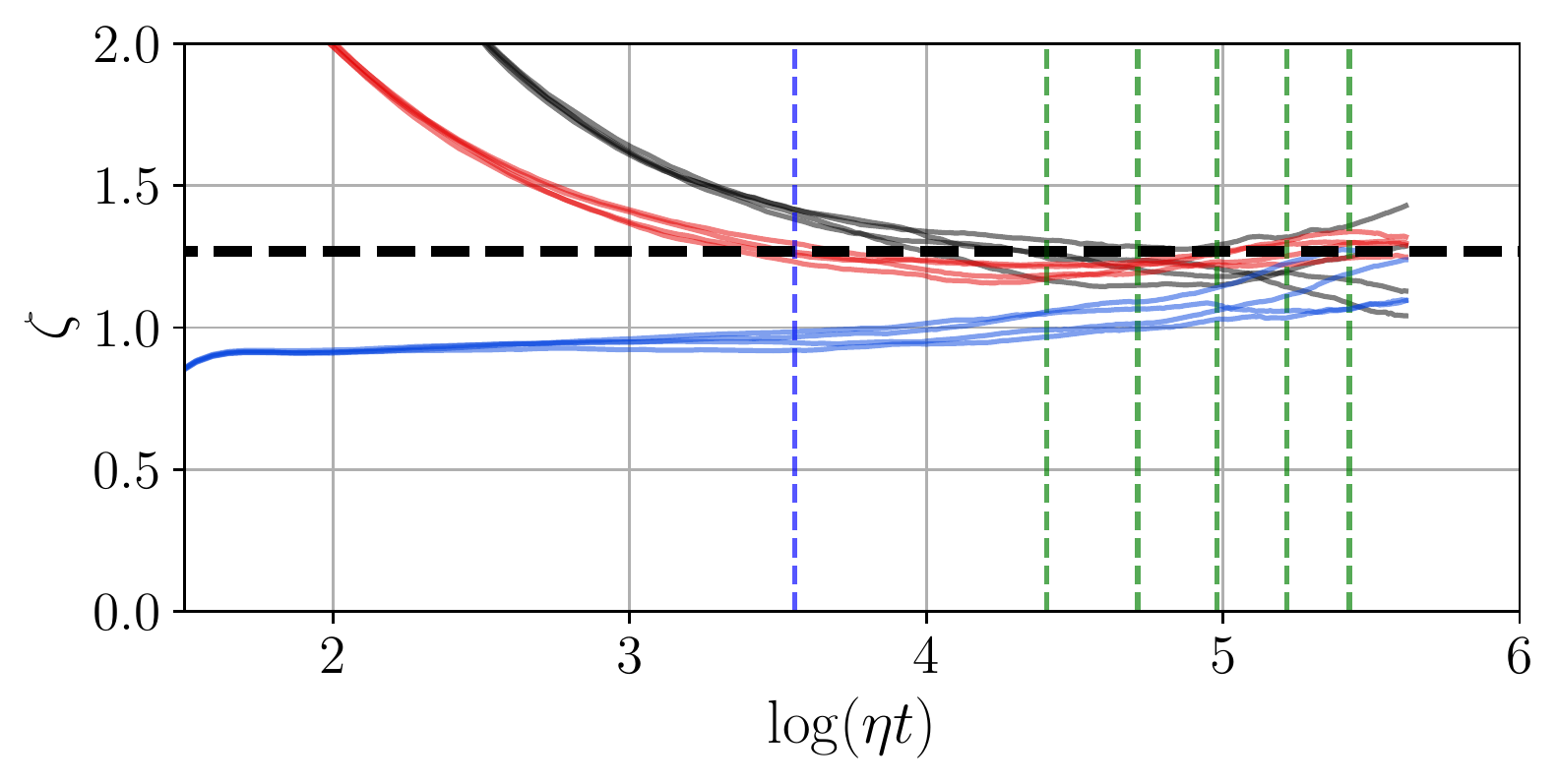}\hfill
\includegraphics[width=0.32\textwidth,clip]{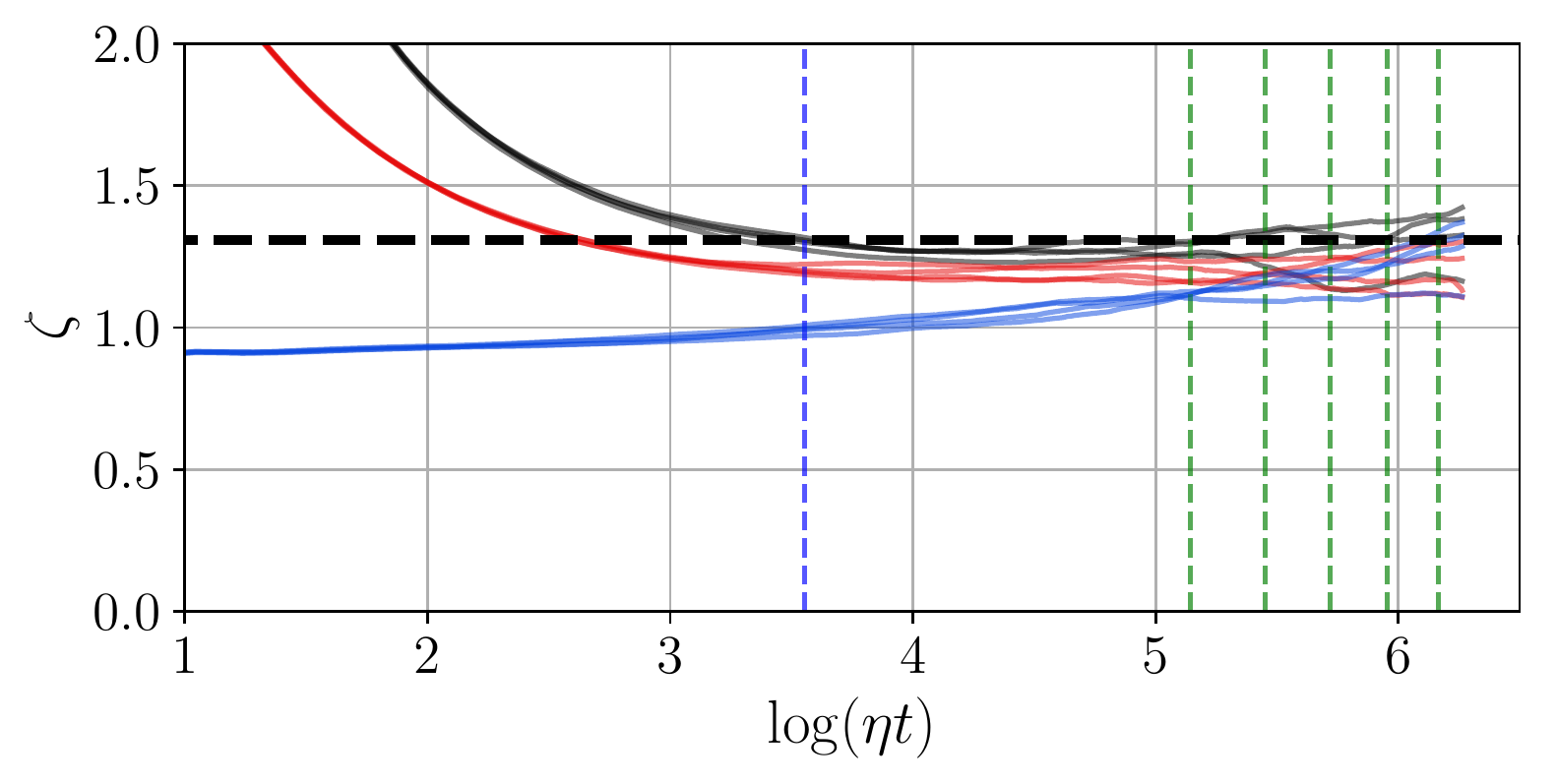}\\\hfill

\includegraphics[width=0.32\textwidth,clip]{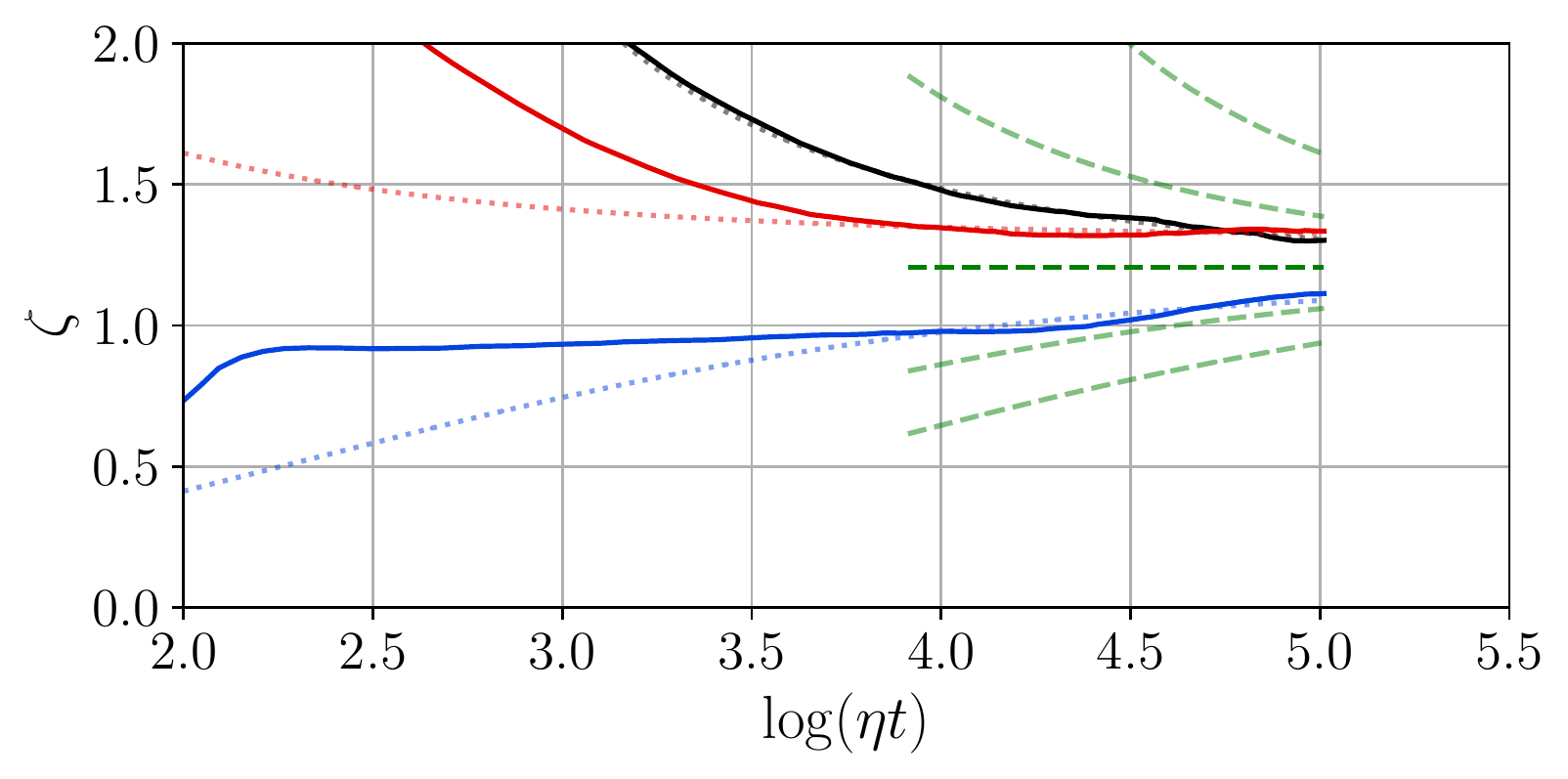}\hfill
\includegraphics[width=0.32\textwidth,clip]{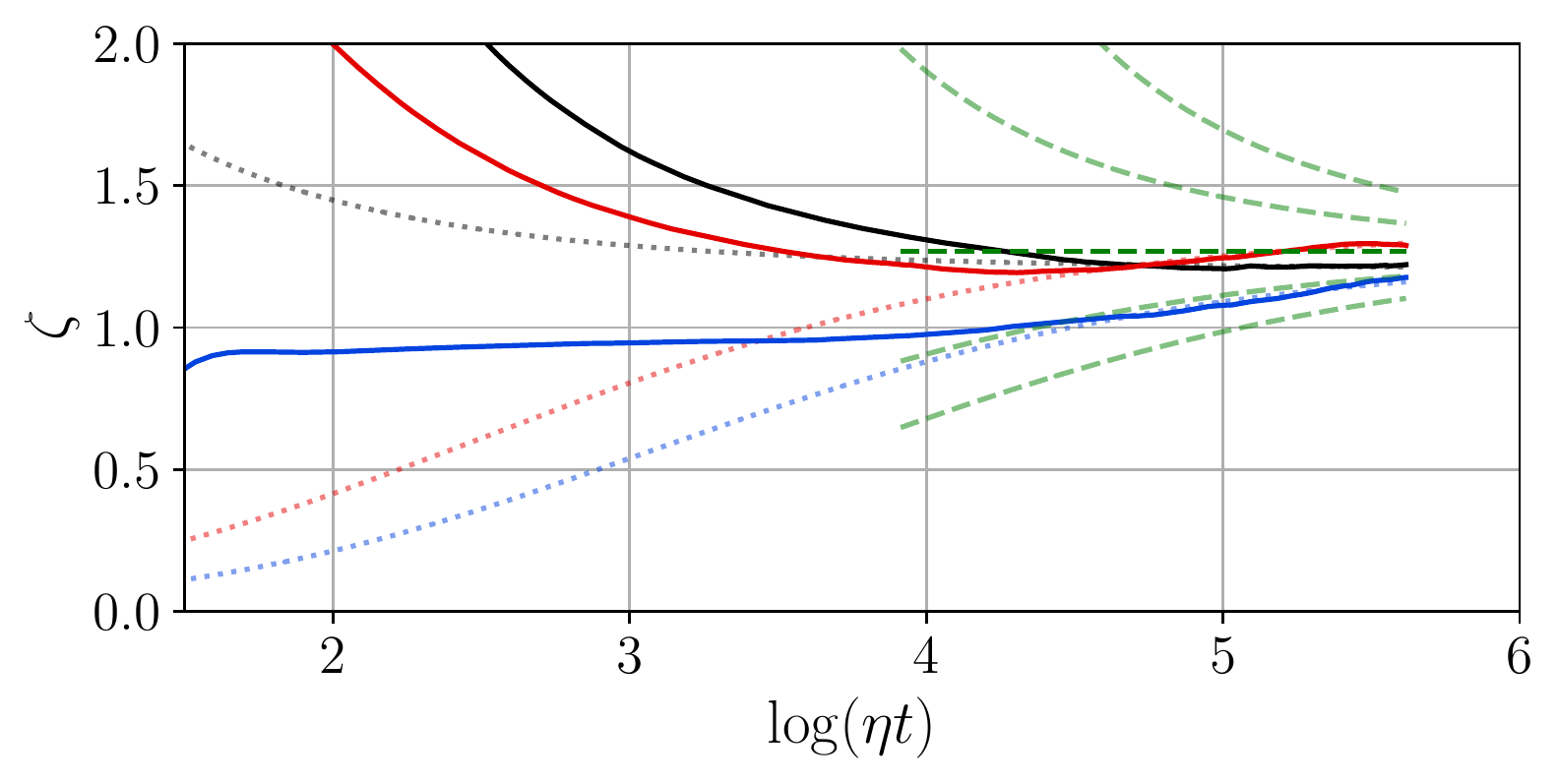}\hfill
\includegraphics[width=0.32\textwidth,clip]{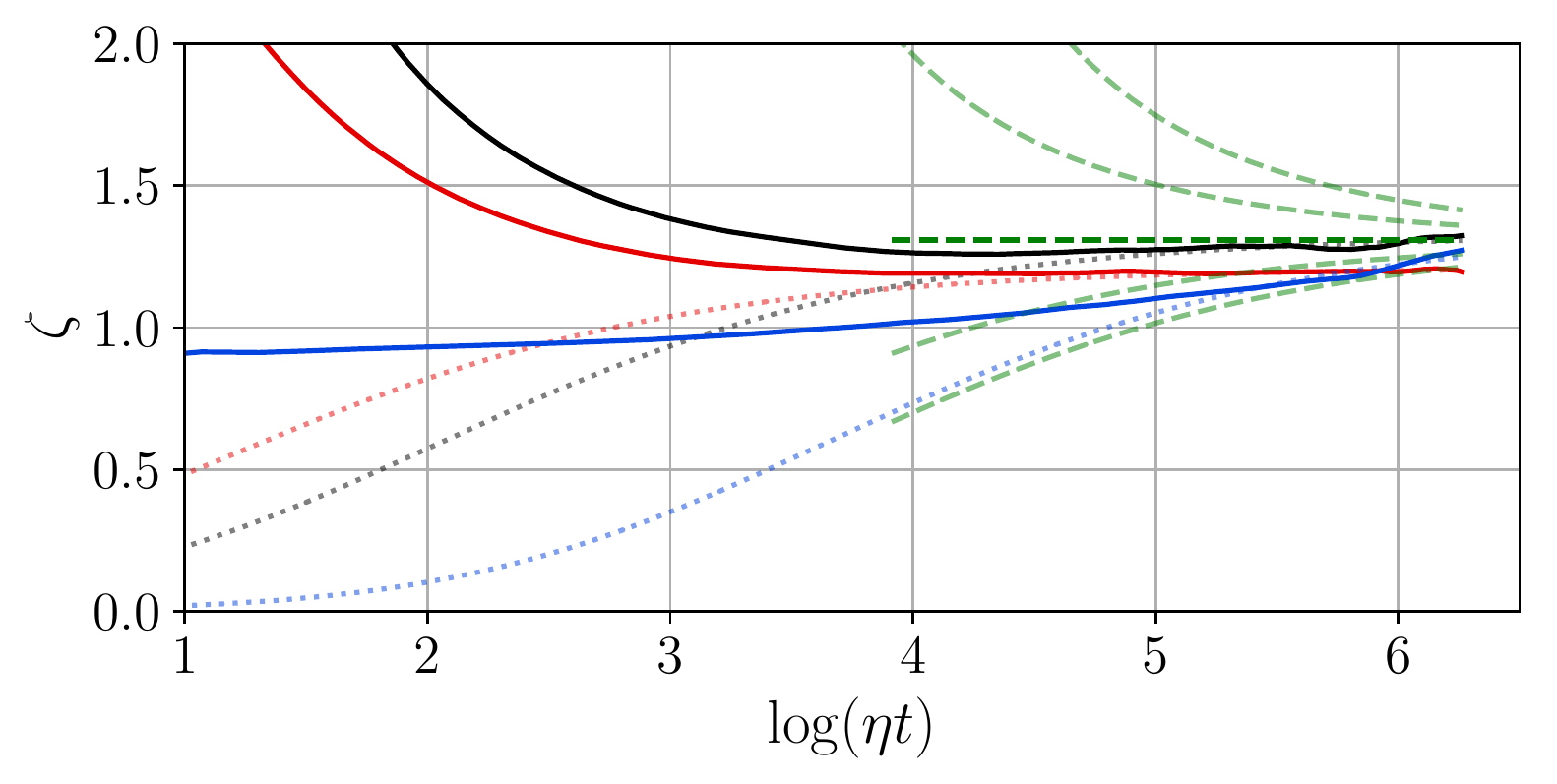}\\\hfill
\caption{
As Fig.~\ref{f:SupMatGrid} but for simulations with constant comoving width.  in this case, the values of $\beta$ for the dashed lines in the figures in the top row, and for $\zeta_0=1/4\beta^2$ in the middle row can be found in Table~\ref{tab:SupMatslopes_errors}.
}
\label{f:SupMatGrids0}
\end{figure}

\begin{figure}[b!]
  \centering
 \includegraphics[width=0.32\textwidth,clip]{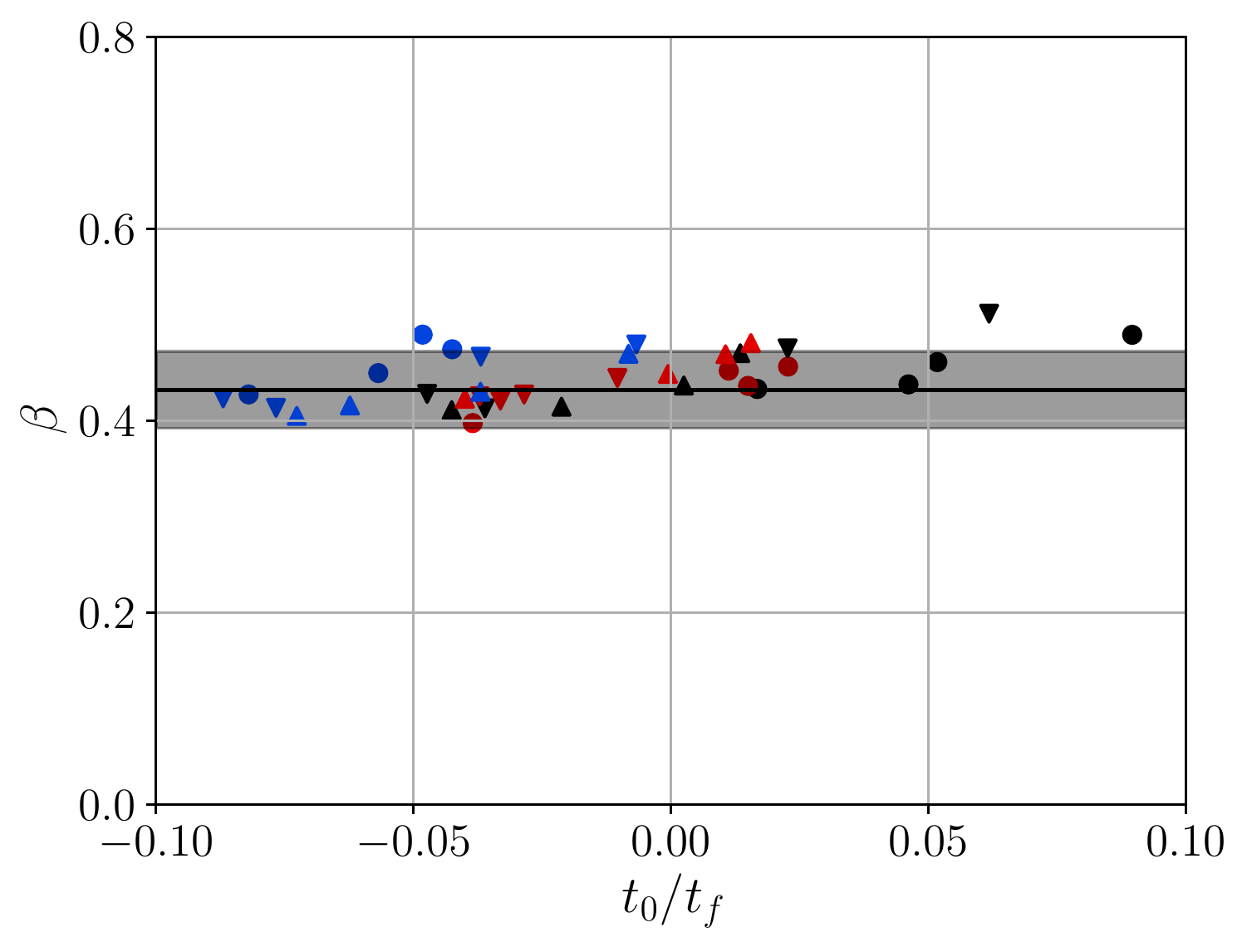}\hfill
\includegraphics[width=0.32\textwidth,clip]{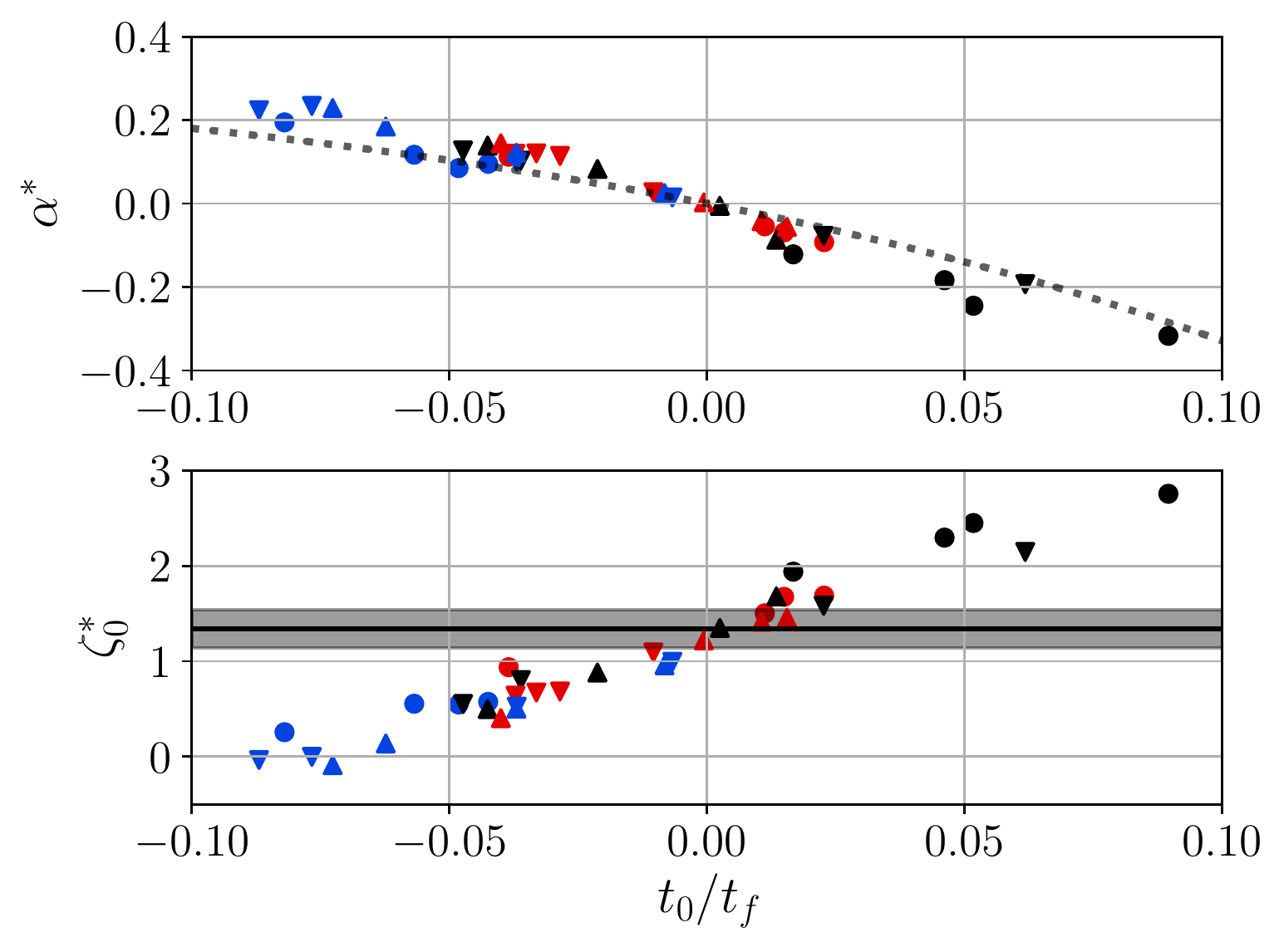}\hfill
\includegraphics[width=0.32\textwidth,clip]{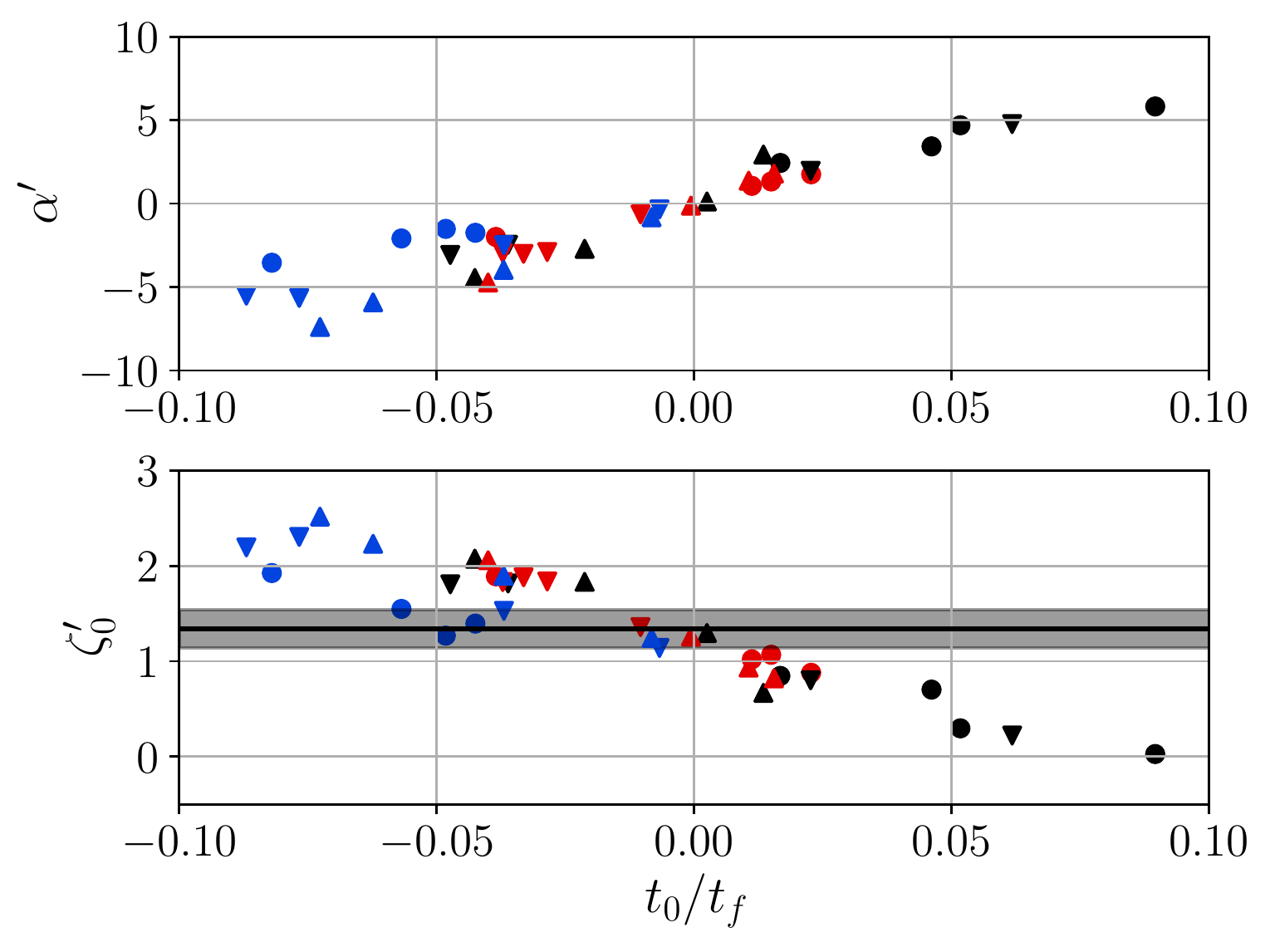}\\\hfill
\caption{
As Fig.~\ref{f:SupMatAlpha} but for simulations with constant comoving width. In this case, the black line with the shaded region in the leftmost figure also represents $\beta=1/2\sqrt{\zeta_0}$ with its 1-$\sigma$ variation, and the  black lines with the shaded region in the bottom of the other two figures represent $\zeta_0$  with its 1-$\sigma$ variation; but the value used is the one in Eq.~\ref{zeta0s0} (i.e., $\zeta_0 = 1.34 \pm 0.22$). 
}
\label{f:SupMatAlphas0}
\end{figure}

Figs.~\ref{f:SupMatGrid}, \ref{f:SupMatAlpha}, \ref{f:SupMatGrids0} and \ref{f:SupMatAlphas0}  support our statements that: 
\begin{enumerate}
\item Simulations which are scaling in the standard sense ($\xi \propto t$) have a slow evolution in the $\zeta$ - $\log(\eta t)$ plane.
\item Simulations converge to $\zeta \simeq 1$ from both above and below.
\item Convergence from below can look like a logarithmic increase in $\zeta$, as observed in 
Refs.~\cite{Gorghetto:2018myk,Kawasaki:2018bzv,Vaquero:2018tib,Buschmann:2019icd}.
\item The coefficients of the logarithm and inverse logarithm have a strong dependence on the ratio $\tOff/t_f$,  
consistent with their being a feature of the initial conditions.
\item The coefficients of the logarithm and inverse logarithm are consistent with zero at $\tOff/t_f=0$.
\item The constant terms in logarithm and inverse logarithm models are consistent with standard scaling at $\tOff/t_f=0$.
\end{enumerate}

\widetext

\end{document}